\documentclass[11pt,a4paper]{article}
\pdfoutput=1

\usepackage{jcappub}

\usepackage{amssymb}
\usepackage{pstricks}
\usepackage{graphicx}
\usepackage{multirow}
\usepackage{hyperref}

\title{Dark Matter interpretation of low energy IceCube MESE excess}

\author{M.~Chianese,}                        
\author{G.~Miele}
\author{and S.~Morisi}

\affiliation{Dipartimento di Fisica {\it Ettore Pancini}, Universit\`a di Napoli {\it Federico II}, and INFN, Sezione di Napoli, Complesso Universitario di Monte S. Angelo, I-80126 Napoli, Italy}

\emailAdd{chianese@na.infn.it}
\emailAdd{miele@na.infn.it}
\emailAdd{stefano.morisi@na.infn.it}

\abstract{The 2-years MESE IceCube events show a slightly excess in the energy range 10--100~TeV with a maximum {\it local} statistical significance of $2.3\sigma$, once a hard astrophysical power-law is assumed. A spectral index smaller than 2.2 is indeed suggested by multi-messenger studies related to $p$--$p$ sources and by the recent IceCube analysis regarding 6-years up-going muon neutrinos. In the present paper, we propose a two-components scenario where the extraterrestrial neutrinos are explained in terms of an astrophysical power-law and a Dark Matter signal. We consider both decaying and annihilating Dark Matter candidates with different final states (quarks and leptons) and different halo density profiles. We perform a likelihood-ratio analysis that provides  a statistical significance up to 3.9$\sigma$ for a Dark Matter interpretation of the IceCube low energy excess.}


\begin{document}

\maketitle

\section{Introduction}

In the last few years IceCube (IC) Neutrino Telescope has been collecting evidences for extraterrestrial neutrinos in the TeV - PeV range~\cite{Aartsen:2013jdh,Aartsen:2014gkd,Aartsen:2014muf,Aartsen:2015knd,Aartsen:2015zva,Aartsen:2016xlq}. Depending on the veto implementation in their analysis, the energy of neutrino events can be as low as 1 TeV for Medium Energy Starting Events (MESE), and 30 TeV for High Energy Starting Events (HESE). Up to now, the origin of such high-energy neutrinos still remains unknown. The pile-up of such events has triggered a huge debate in the scientific community about their possible astrophysical origin, as well as an intriguing connection with Dark Matter (DM). In particular IC events in the PeV region have been related to the decay of very heavy DM particles \cite{Anisimov:2008gg,Feldstein:2013kka,Esmaili:2013gha,Bai:2013nga,Ema:2013nda,Esmaili:2014rma,Bhattacharya:2014vwa, Higaki:2014dwa,Bhattacharya:2014yha,Rott:2014kfa,Ema:2014ufa, Murase:2015gea,Dudas:2014bca,Fong:2014bsa,Boucenna:2015tra,Aisati:2015vma,Ko:2015nma,Dev:2016qbd,Fiorentin:2016avj,DiBari:2016guw}.

Different astrophysical sources have been proposed in literature as potential candidates providing a contribution to the extraterrestrial neutrino flux either in a particular energy range or in the whole TeV - PeV range. Among such sources one can quote the stellar remnants in star-forming galaxies like extragalactic Supernovae and Hypernovae remnants~\cite{Chakraborty:2015sta}, active galactic nuclei like blazars~\cite{Stecker:1991vm,Kalashev:2014vya}, and gamma-ray bursts~\cite{Waxman:1997ti}. In general, since it is commonly believed that observed astrophysical neutrinos have some correlation with hadronic cosmic-rays, under such hypothesis the neutrino energy spectrum has to show a power-law behavior, $E_\nu^{-\gamma}$, where $\gamma$ is the spectral index. In first approximation, Fermi acceleration mechanism at shock fronts predicts a spectral index $\gamma=2.0$, but depending on the particular neutrino production mechanism there can be deviations from such a value. For instance, if neutrinos arise from hadronuclear $p$-$p$ interactions then $\gamma \lesssim 2.2$~\cite{Loeb:2006tw,Kelner:2006tc}, whereas if neutrino arises from photohadronic $p$-$\gamma$ interactions we have $\gamma \gtrsim 2.3$~\cite{Winter:2013cla}.

Moreover, it is also well-known that production mechanisms for neutrinos give rise at the source (or nearby) to gamma-rays as well. The relation between the energy spectrum of neutrinos and gamma-rays is in general model dependent. Due to the loss of energy through inverse Compton and pair production, the observed photons have a degraded energy typically below TeV, and therefore can be observed by Fermi-LAT~\cite{Ackermann:2014usa}. Hence a multi-messenger analysis is able to provide quite strong constraints on the features of the neutrino flux produced in astrophysical environments. For instance, in Ref.s~\cite{Murase:2013rfa,Bechtol:2015uqb,Chakraborty:2016mvc} it has been shown that, in case of $p$-$p$ astrophysical sources, the Fermi-LAT experimental bounds on the gamma-rays spectrum constrain the spectral index to be smaller than 2.2 (see also Ref.~\cite{Murase:2016gly}). Moreover, in Ref.~\cite{Bechtol:2015uqb} it is pointed out that models considering star-forming galaxies as dominant neutrino sources are in tension with data, since such sources can provide at most a contribution to the neutrino flux of $\sim 30\%$ at 100~ TeV and $\sim60\%$ at 1~PeV. However, very recently, Ref.~\cite{Chakraborty:2016mvc} claimed that this tension can be reconciled by considering the uncertainties on the gamma-rays absorption in the astrophysical environment itself and in the intergalactic medium. Concerning the $p$-$\gamma$ sources, the searches for spatial and temporal correlation with the Fermi-LAT observations point out that gamma-ray bursts~\cite{Aartsen:2014aqy} and blazars~\cite{Glusenkamp:2015jca,Schimp:2015xha} can account only for $\sim1\%$ and $\sim20\%$ of the IceCube diffuse neutrino spectrum, respectively. Nevertheless, it is worth observing that the multi-messenger constraints do not hold in case of astrophysical sources where gamma-rays are absorbed in the surrounding environment, leaving room for hidden $p$-$\gamma$ sources~\cite{Murase:2015xka}. For instance, in case of the low-luminosity gamma-ray bursts with choked jets~\cite{Senno:2015tsn} (and references therein), no production of detectable gamma-rays in the Fermi-LAT energy range (GeV--TeV) is expected. 

In view of the previous considerations and assuming a subdominant contribution from hidden (opaque) $p$-$\gamma$ astrophysical sources, one can regard as fully compatible models of neutrino flux a spectral index between 2.0 (Fermi acceleration mechanism) and about 2.2 (fit of Fermi-LAT gamma-rays from hadronuclear reactions).

By analyzing the IceCube results in terms of a single power-law, the data seem to prefer a steep neutrino flux, behaving as $E_\nu^{-2.46}$ in case of 2-years MESE data~\cite{Aartsen:2014muf}, and $E_\nu^{-2.58}$ in case of 4-years HESE ones~\cite{Aartsen:2015zva}. As already stated, such soft exponents can be hardly reconciled with acceleration and production mechanisms, and result in tension with the multi-messenger constraints. On the other hand, the latest IceCube analysis on the 6-years up-going muon neutrinos~\cite{Aartsen:2016xlq} provides a spectral index $\gamma=2.13$ as best-fit, implying a significant tension ($3.3\sigma$) with respect to the combined analysis of the previous data samples ~\cite{Aartsen:2015knd}. These results lead to the conclusion that an additional component to the neutrino flux may be present at low energy ($E_\nu\leq100$~TeV), pointing towards a {\it multi-components} scenario, and that the origin of such a new contribution may be galactic~\cite{Aartsen:2016xlq}.

Very recently, in Ref.~\cite{Chianese:2016opp} particular attention has been payed to the possible presence of a $2\sigma$ bump in the 4-years HESE data in the energy range 60 -- 100~TeV. Such an {\it excess} is indeed shown in the number of neutrino events once both the background (conventional and prompt atmospheric neutrinos and muons) and an astrophysical power-law with $\gamma=2.0$ (being considered as benchmark prediction) have been subtracted. In Ref.~\cite{Chianese:2016opp} such a multi-component hypothesis is scrutinized by using the spatial information about the arrival direction of neutrino events, comparing an extra component of astrophysical origin with a more intriguing DM one (for other spatial studies of IC events see for instance Ref.s~\cite{{Esmaili:2014rma,Pagliaroli:2016lgg,Troitsky:2015cnk}}). The possible sources of the above excess (astrophysical vs DM) are therefore characterized by analyzing their statistical relevance and making a forecast for future Neutrino Telescopes.

\begin{figure}[h]
\centering
\includegraphics[width=0.45\textwidth]{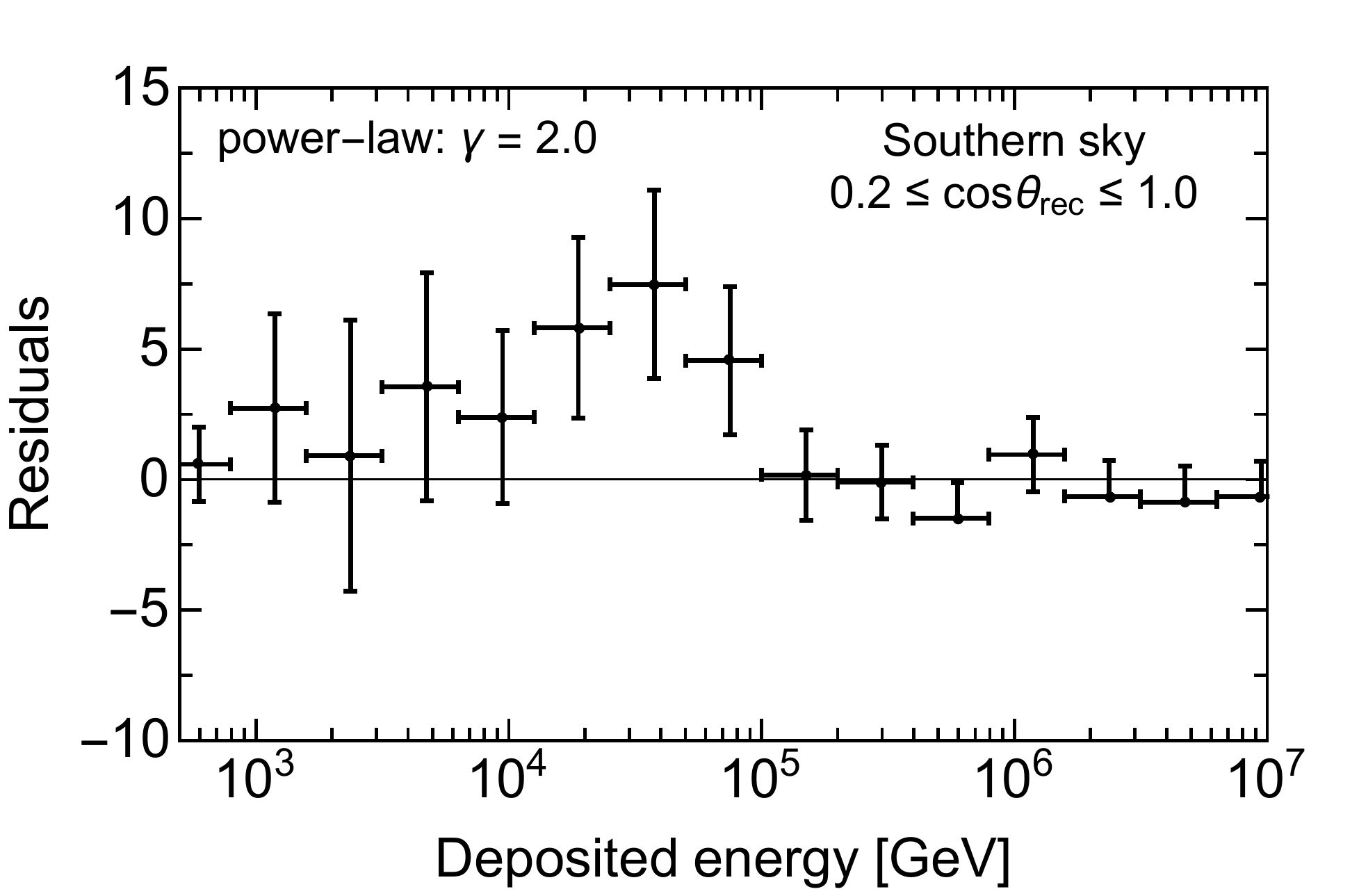}
\hskip3.mm
\includegraphics[width=0.45\textwidth]{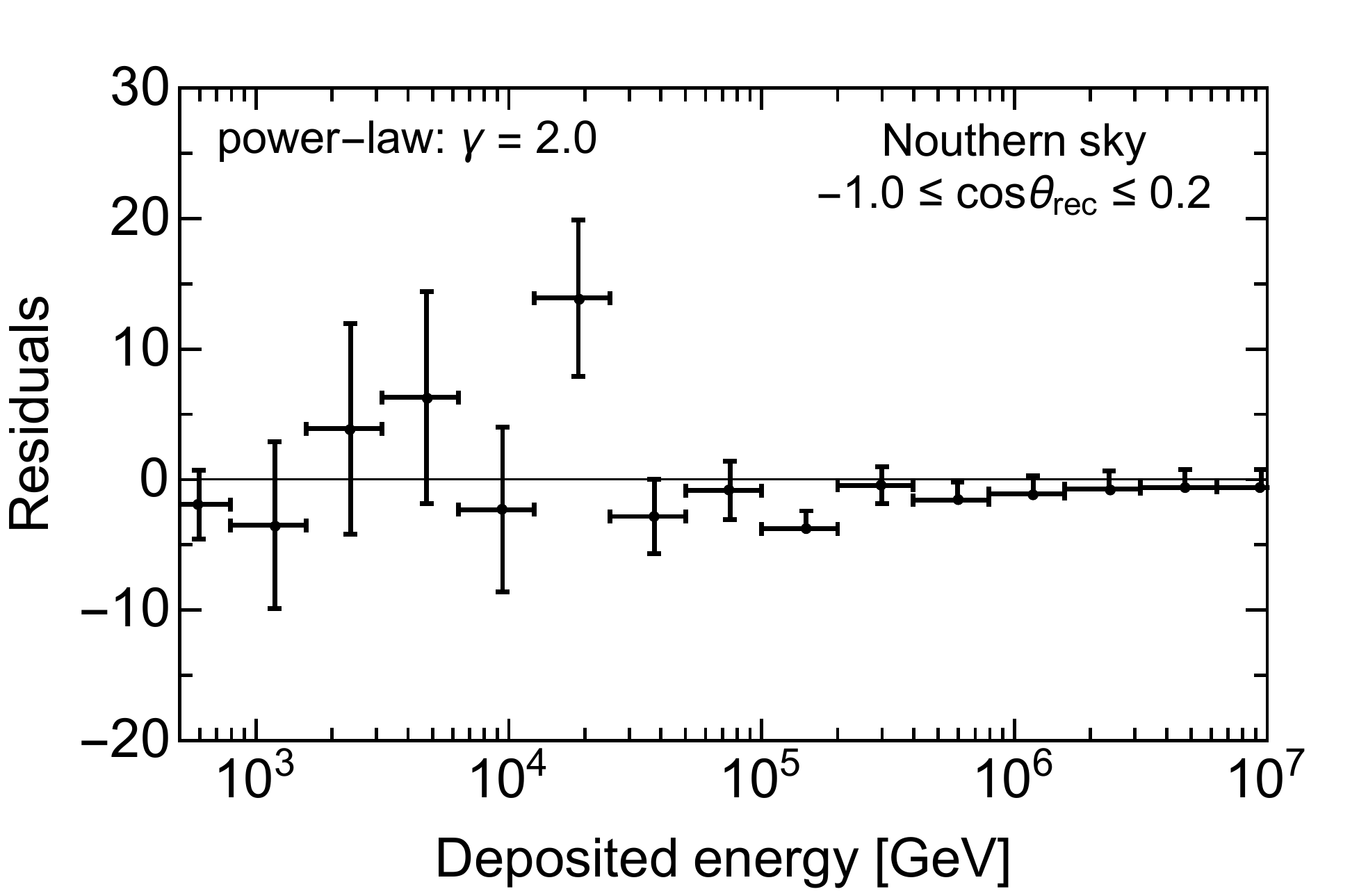}
\includegraphics[width=0.45\textwidth]{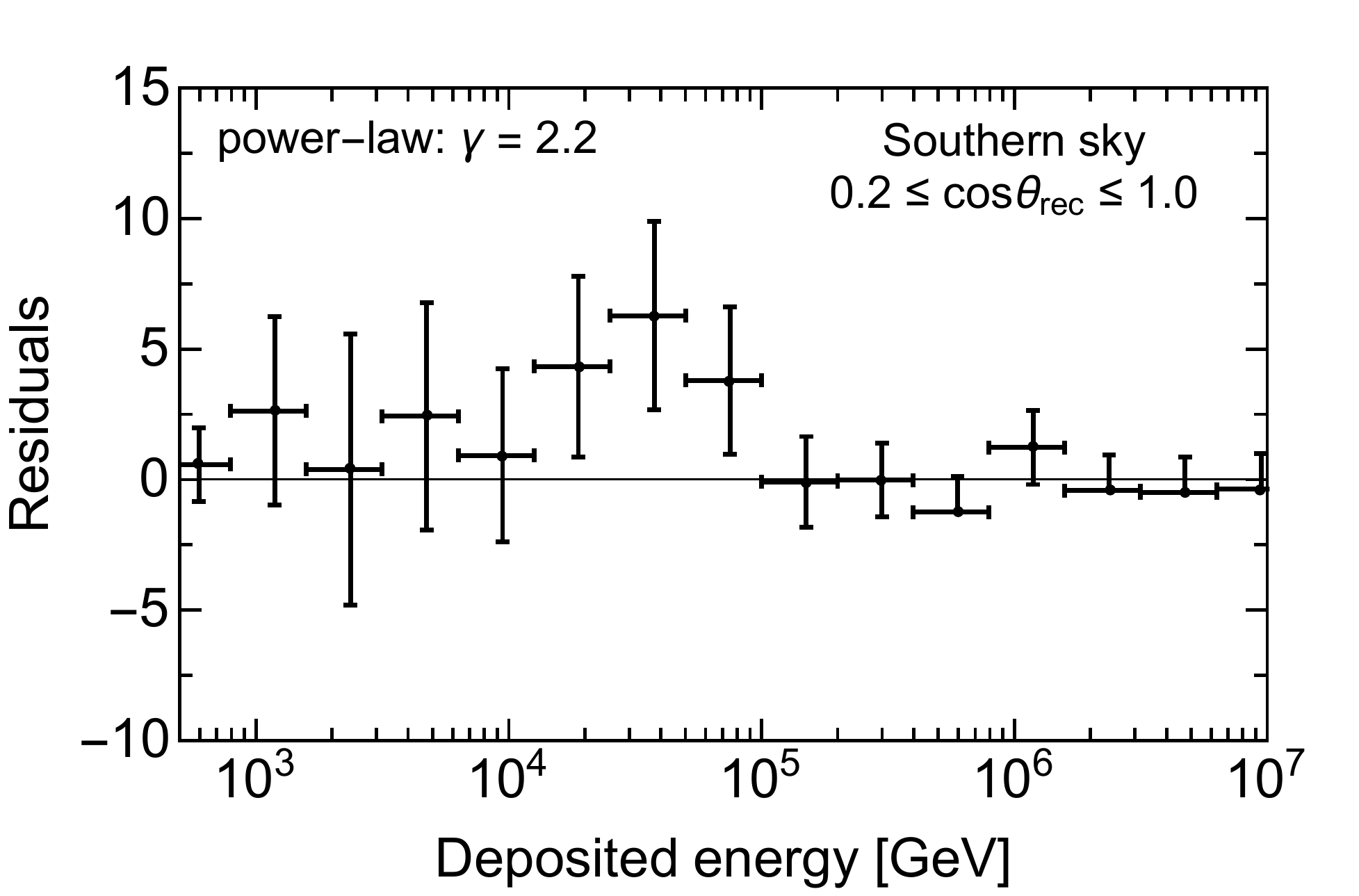}
\hskip3.mm
\includegraphics[width=0.45\textwidth]{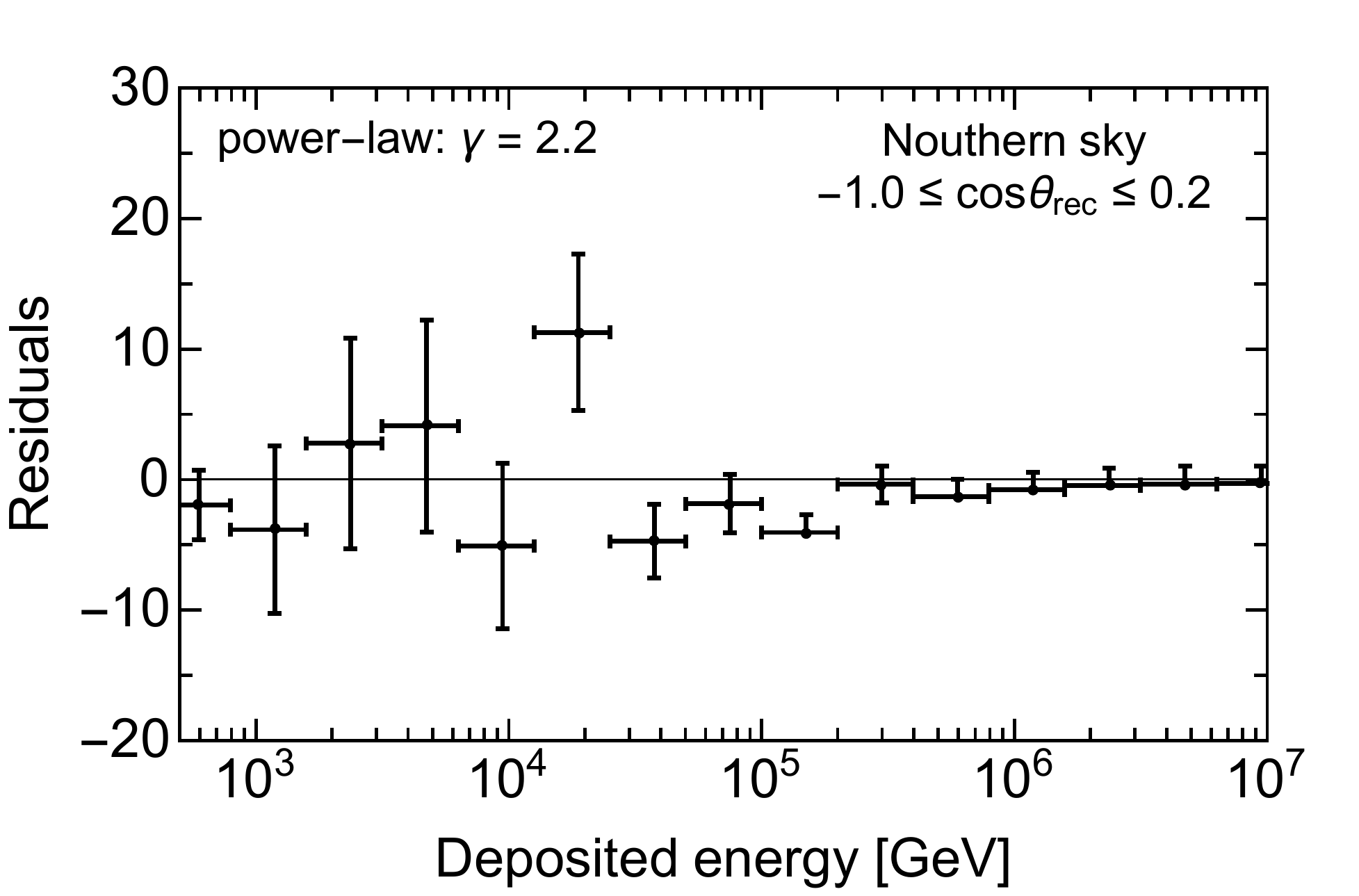}
\caption{\label{fig:residual}Residuals in the number of neutrino events as a function of the neutrino deposited energy in the southern and northern hemispheres, once a single astrophysical power-law with spectral index 2.0 and 2.2 is considered.}
\end{figure}

In the present paper, inspired by the approach of Ref.~\cite{Chianese:2016opp}, we analyze the 2-years MESE data in order to  provide statistical evidence to a possible explanation of the low energy excess in terms of decaying/annihilating DM neutrino flux in addition to a hard astrophysical power-law ($p$-$p$ sources). The MESE data sample is indeed more suitable for such an analysis at low energy since the different veto implementation provides a larger sensitivity (larger statistics) for $E_\nu \lesssim 100$~TeV. In particular, assuming the IceCube best-fit spectral index of 2.46, the 2-years MESE events show an excess in the energy range 10--100~TeV with a maximum {\it local} statistical significance of $1.5\sigma$ in both southern and northern hemispheres. However, in case of a harder spectrum the maximum {\it local} deviation of the excess further increases towards $2.3\sigma$ and $1.9\sigma$ level for spectral index 2.0 and 2.2, respectively. This result can be seen in Fig.~\ref{fig:residual} where the residuals in the number of neutrino events are reported once the atmospheric background and a single astrophysical power-law are subtracted by the data. Remarkably, in view of the previous considerations, if the observed extraterrestrial neutrinos have hadronuclear $p$-$p$ astrophysical origin, according to the gamma-ray constraints, the presence of some {\it extra} neutrino component below 100~TeV would be suggested by IC data.

Hence, in the present analysis, for the extraterrestrial neutrinos we assume a two-components scenario where the astrophysical power-law has a spectral index belonging to the interval $\left[2.0,2.2\right]$ and the low energy excess is related to decaying/annihilating DM particles. Hereafter, we will consider only the extreme values 2.0 and 2.2 that are representative of the whole interval for $p$--$p$ astrophysical sources. In order to quantity how the fit is statistically improved by adding a DM neutrino component, we perform a likelihood-ratio statistical analysis on the neutrino energy spectrum splitting the data into southern and northern hemispheres\footnote{It would be interesting to perform a study restricted to the Galactic Center, but this would require the detailed knowledge of the IC effective area as a function of the reconstructed arrival direction that is not public.}. To be almost model independent, we make two different and quite extreme assumptions for the DM halo density profile, namely the Navarro-Frenk-White and Isothermal distributions. Moreover, we consider different decaying/annihilating channels both in leptons and quarks, since they provide different features for the neutrino flux. Furthermore, we study also the peculiar case where the neutrinos are produced by three-bodies decays of leptophilic DM particles, model that has already been proposed to explain PeV neutrinos in Ref.~\cite{Boucenna:2015tra}.

The paper is organize as follows. In Section 2 we review the two-components scenario giving details on the neutrino flux expected for decaying/annihilating DM for different final states. In Section 3 we discuss the gamma-ray constraints on DM models. In Section 4 we describe the statistical analysis and report the results. Finally, Section 5 contains our conclusions.

\section{Neutrino flux in the two-components scenario}

As already stated, we assume that the neutrino flux observed by IceCube is explained in terms of atmospheric conventional neutrinos, penetrating muons and two additional components, one originated by astrophysical sources and one given either by decaying or annihilating DM (see Ref.~\cite{Chen:2014gxa} for a two-component scenario with two astrophysical power-laws). The contribution of atmospheric prompt neutrinos produced by the decays of charm mesons is not considered since, as recently shown in Ref.s~ \cite{Bhattacharya:2015jpa,Gauld:2015kvh,Halzen:2016thi,Bhattacharya:2016jce}, it results to be almost negligible. Under these assumptions, the total differential flux of extraterrestrial neutrinos at the Earth, at a given energy $E_\nu$, per flavour $\alpha$ and per unit solid angle $\Omega$, is given by
\begin{equation}
\frac{{\rm d}\phi_\alpha}{{\rm d}E_\nu {\rm d}\Omega} = \frac{{\rm d}\phi^{\rm Astro}_\alpha}{{\rm d}E_\nu {\rm d}\Omega} + \frac{{\rm d}\phi^{\rm DM}_\alpha}{{\rm d}E_\nu {\rm d}\Omega}\,,
\label{eq:tot_flux}
\end{equation}
where the first term (Astro) is the astrophysical neutrino component, while the second term is the DM one.

In general, the astrophysical component is parametrized at the Earth by a power-law whose expression is given by
\begin{equation}
\frac{{\rm d}\phi^{\rm Astro}_\alpha}{{\rm d}E_\nu {\rm d}\Omega} = \phi^{\rm Astro}_0 \left( \frac{E_\nu}{100~{\rm TeV}} \right)^{-\gamma}\,,
\label{eq:astro}
\end{equation}
where $\phi^{\rm Astro}_0$ represents the normalization of the neutrino flux at 100~TeV and $\gamma$ is the spectral index. This parametrization does not depend on the angular coordinates and provides an isotropic flux as expected for extragalactic astrophysical sources. Moreover, we assume an equal flux for each neutrino flavour~$\alpha$. Indeed, in standard astrophysical sources, a flavour ratio 1:2:0 is expected at the source and, once the neutrino propagation and oscillations are taken into account, it becomes 1:1:1 at the Earth (see Ref.s~\cite{Palladino:2015zua,Arguelles:2015dca,Bustamante:2015waa} for studies on different flavour ratios and Ref.~\cite{Aartsen:2015ivb} for the corresponding measurements of the IC Collaboration).

As discussed in the previous Section, we consider two extreme values for the spectral index $\gamma$, representing fully compatible astrophysical models:
\begin{itemize}
\item spectral index $\gamma = 2.0$, i.e. the benchmark scenario provided by the standard Fermi acceleration mechanism at shock fronts;
\item spectral index $\gamma = 2.2$, which corresponds to the upper bound for $p$--$p$ astrophysical sources provided by the Fermi-LAT gamma-ray spectrum in the GeV--TeV energy~\cite{Murase:2013rfa}.
\end{itemize}
For each choice of the spectral index, the flux normalization $\phi^{\rm Astro}_0$ is then obtained by fitting its value through a maximum-likelihood procedure.

In the following, we will discuss the neutrino flux produced by decaying/annihilating DM particles.

\subsection{Neutrino Flux from Dark Matter}

Neutrinos can be produced as primary or secondary particles by the decays of an unstable DM particle or by the pair-annihilation of stable DM particles. The DM differential neutrino flux of a flavour~$\alpha$ at the Earth is composed by two contributions
\begin{equation}
\frac{{\rm d}\phi^{\rm DM}_\alpha}{{\rm d}E_\nu {\rm d}\Omega} = \sum_{\beta} P_{\alpha\beta} \left[ \frac{{\rm d}\phi^{\rm G}_\beta}{{\rm d}E_\nu {\rm d}\Omega} + \frac{{\rm d}\phi^{\rm EG}_\beta}{{\rm d}E_\nu {\rm d}\Omega} \right]\,,
\label{eq:flux}
\end{equation}
where the first term refers to the Galactic (G) contribution of the Milky Way, while the second one represents the Extragalactic (EG) component. In this case, since the flavour ratio at the source depends on the particular Dark Matter model considered as well as on the energy, we take into account the neutrino flavour oscillations during the propagation, which are represented by the mixing probabilities $P_{\alpha\beta}$ that a neutrino of flavour $\beta$ is converted into a neutrino of flavour $\alpha$. For long baseline oscillations, the mixing probabilities are equal to 
\begin{equation}
\begin{array}{lclcl}
P_{ee}=0.573\,, & \qquad & P_{e\mu}=0.348\,, & \qquad & P_{e\tau} = 0.150\,, \\
&&&&\\
P_{\mu\mu}=0.348\,, & \qquad & P_{\mu\tau}=0.375\,, & \qquad & P_{\tau\tau} = 0.475\,,
\end{array}
\end{equation}
where the values have been taken from Ref.~\cite{Aisati:2015vma}. In case of decaying (dec.) DM particles, the Galactic and the Extragalactic fluxes in Eq.~\eqref{eq:flux} have, respectively, the expressions
\begin{eqnarray}
\left. \frac{{\rm d}\phi^{\rm G}_\beta}{{\rm d}E_\nu {\rm d}\Omega} \right|_{\rm dec.} & = & \frac{1}{4\pi \, m_{\rm DM} \, \tau_{\rm DM}} \frac{{\rm d}N_\beta}{{\rm d}E_\nu} \int_0^\infty ds \, \rho_h\left[r\left(s,\ell,b\right)\right]\,,\\
\left. \frac{{\rm d}\phi^{\rm EG}_\beta}{{\rm d}E_\nu {\rm d}\Omega} \right|_{\rm dec.} & = &  \frac{\Omega_{\rm DM}\rho_c}{4\pi \, m_{\rm DM} \, \tau_{\rm DM}} \int_0^\infty dz \,  \frac{1}{H\left(z\right)}\left.\frac{{\rm d}N_\beta}{{\rm d}E_\nu}\right|_{E'=E\left(1+z\right)}\,.
\end{eqnarray}
Here, the quantities $m_{\rm DM}$ and $\tau_{\rm DM}$ denote the mass and the lifetime of DM particles, respectively. The Galactic term is proportional to the integral over the line-of-sight $s$ of the galactic DM halo density $\rho_h\left(r\right)$ where $r = \sqrt{s^2+R^2-2sR\cos\ell\cos b}$ with $R = 8.5$~kpc and $\left(b,\ell\right)$ being the Galactic coordinates. The EG flux is instead obtained by integrating over the redshift $z$ (the absorption of neutrinos in the intergalactic medium is negligible). Moreover, in the second expression $\rho_c = 5.5 \times 10^{-6}\,{\rm GeV \, cm}^{-3}$ is the critical energy density, $H\left(z\right)=H_0\sqrt{\Omega_\Lambda+\Omega_m\left(1+z\right)^3}$ is the Hubble expansion rate with $h = H_0/100\,{\rm km\,s^{-1}\,Mpc^{-1}}=0.6711$, $\Omega_{\rm DM} = 0.2685$, $\Omega_\Lambda = 0.6825$ and $\Omega_m = 0.3175$ according to Planck analysis~\cite{Ade:2015xua}. Finally, the quantity ${\rm d}N_\beta/{\rm d}E_\nu$ is the energy spectrum of $\beta$-flavour neutrinos produced by DM particles. This quantity depends on the particular DM interaction with the SM particles and, in general, is obtained by means of a Monte Carlo procedure.

In case of annihilating (ann.) DM particles instead, the two contributions to the neutrino flux of Eq.~\eqref{eq:flux} are equal to
\begin{eqnarray}
\left. \frac{{\rm d}J^{\rm G}_\beta}{{\rm d}E_\nu {\rm d}\Omega} \right|_{\rm ann.} & = & \frac12 \frac{\left< \sigma v \right>}{4\pi \, m_{\rm DM}^2} \frac{{\rm d}N_\beta}{{\rm d}E_\nu} \int_0^\infty ds \, \rho_h^2\left[r\left(s,\ell,b\right)\right]\,,\\
\left. \frac{{\rm d}J^{\rm EG}_\beta}{{\rm d}E_\nu {\rm d}\Omega} \right|_{\rm ann.} & = & \frac12 \frac{\left< \sigma v \right> \, \left(\Omega_{\rm DM}\rho_c\right)^2}{4\pi \, m_{\rm DM}^2} \int_0^\infty dz \, \frac{B\left(z\right) \, \left(1+z\right)^3}{H\left(z\right)}\left.\frac{{\rm d}N_\beta}{{\rm d}E_\nu}\right|_{E'=E\left(1+z\right)}\,.
\end{eqnarray}
In the previous expressions, the quantity $\left< \sigma v \right>$ is the thermally averaged cross section and $B\left(z\right)$ is the {\it boost factor} (or {\it clumpiness factor}), which parametrizes the effect of the inhomogeneities of the DM distribution in the intergalactic medium. In the present study, we adopt the cosmological boost factor reported in Ref.~\cite{Cirelli:2010xx}, which is obtained by considering a Navarro-Frenk-White distribution in each sub-halo and the {\it power-law} model~\cite{Neto:2007vq,Maccio':2008xb} with a minimum halo mass of $10^{-6}M_\odot$~\cite{Martinez:2009jh,Bringmann:2009vf} for the concentration parameter. However, it is worth observing that the boost factor is affected by large uncertainties, and the model adopted in this analysis has to be considered as a benchmark model. Indeed, different models for the concentration parameter, as well as different DM distributions, can be considered. This implies an uncertainty of orders of magnitude for the cosmological boost factor at low redshift. Such large uncertainty mainly affects the angular distribution of neutrino arrival directions, as shown in Ref.~\cite{Chianese:2016opp}.

In the following, we consider two DM halo density profiles that provide two extreme cases (predicting different angular distributions of neutrino arrival directions): the Navarro-Frenk-White distribution (NFW) and the Isothermal one (ISO). The NFW distribution is given by
\begin{equation}
\rho^{\rm NFW}_h \left(r\right)= \frac{\rho_s}{r/r_s\left(1+r/r_s\right)^2}\,,
\label{eq:NFW}
\end{equation}
with $r_s = 24.42$~kpc and $\rho_s = 0.184 \, {\rm GeV \, cm}^{-3}$, while the ISO distribution results to be
\begin{equation}
\rho^{\rm ISO}_h \left(r\right)= \frac{\rho_s}{1+\left(r/r_s\right)^2}\,,
\label{eq:ISO}
\end{equation}
with $r_s = 4.38$~kpc and $\rho_s = 1.879 \,{\rm GeV \, cm}^{-3}$ (see Ref.~\cite{Cirelli:2010xx} and references therein). In particular, the former is enhanced towards the Galactic Center, while the latter is practically uniform.

In the present paper, we examine different decay/annihilation channels of DM particles (hereafter denoted as $\chi$) into SM particles providing different shape of the neutrino energy spectrum ${\rm d}N/{\rm d}E_\nu$ in order to cover all the possible phenomenological scenarios. In particular we consider:
\begin{itemize}
\item $\boldsymbol{\chi \to f\overline{f}}$: a bosonic DM particle $\chi$ decaying into a couple of SM fermions, like quarks ($\chi \to t\overline{t}$ or $\chi \to b\overline{b}$) and leptons ($\chi\to \mu^-\mu^+$ or $\chi\to\tau^-\tau^+$). In this case, we use the neutrino energy spectra reported in Ref.~\cite{Cirelli:2010xx}.

\item $\boldsymbol{\chi \to \ell^-\ell^+\nu}$: a fermionic DM particle $\chi$ producing neutrinos by leptophilic three-bodies decays. This scenario has two intriguing characteristics, i.e. the neutrino spectrum is spread differently from the case of two-bodies decays and it is peaked in a particular energy range due to the absence of quarks in the final states. The leptophilic three-bodies decay comes from the non-renormalizable effective operator
\begin{equation}
\mathcal{L} \supset \frac{1}{\Lambda^2}  \overline{L}\ell_{\rm R}\overline{L}\chi \,,
\label{eq:A4op}
\end{equation}
where $\Lambda$ represents the energy scale of new physics, and $L$ and $\ell_{\rm R}$ stand for the left-handed doublet and right-handed singlet of the SM leptonic sector, respectively. Such an operator arises from discrete non-Abelian flavour symmetries that are introduced to address for the flavour problem. In particular, assuming the well-known $A_4$ symmetry~\cite{Babu:2002dz,Altarelli:2005yx}, only the effective operator of Eq.~\eqref{eq:A4op} is allowed~\cite{Haba:2010ag}. In this case, the operator has a cyclic flavour structure, like $\overline{e_{\rm L}}\mu_{\rm R}\overline{\nu_\tau}\chi$ plus cyclic permutations of flavour indexes. The energy neutrino spectrum has been evaluated by using a Monte Carlo procedure that takes also into account the production of secondary neutrinos. This leptophilic model has already been proposed in Ref.~\cite{Boucenna:2015tra} as a candidate able to explain the PeV neutrinos reported in the 4-years HESE data set. Moreover, in Ref.~\cite{Chianese:2016smc} it is shown that the same operator can account also for the DM production through a freeze-in mechanism, providing a minimal extension of the SM.

\item $\boldsymbol{\chi \chi \to f \overline{f}}$: two fermionic DM particles annihilating into SM particles $f=b,t,\mu,\tau$ (only one channel at a time is considered). As for the decaying case, we use the tables of Ref.~\cite{Cirelli:2010xx} for the neutrino energy spectra.
\end{itemize}

\vspace{3.mm}

The energy spectra provided in Ref.~\cite{Cirelli:2010xx} have been evaluated up to a DM mass of 100~TeV by means of a Monte Carlo procedure that takes into account the decays of unstable SM particles (like pions) and the electroweak radiative corrections, which are relevant for heavy DM particles. In order to perform the analysis for DM masses larger than 100~TeV, we extrapolate such spectra up to a DM mass of 10 PeV by considering an appropriate rescaling. In particular, for $m_{\rm DM} \geq 100$~TeV, the energy spectrum is given by
\begin{equation}
\left.\frac{{\rm d}N}{{\rm d}E}\right|_{m_{\rm DM}\geq100\,{\rm TeV}} = \frac{1}{E \ln(10)} \left.\frac{{\rm d}N}{{\rm d}\log x}\left(x\right)\right|_{100~{\rm TeV}}
\end{equation}
where the quantity ${\rm d}N/{\rm d}\log x$, provided in Ref.~\cite{Cirelli:2010xx}, is a function of the variable $x=E/m_{\rm DM}$ and it is evaluated with a DM mass of 100~TeV. As will be shown in Section~4, such an approximation does not strongly affect the most relevant DM scenarios that are in correspondence of a $\mathcal{O}\left(100~\text{TeV}\right)$ DM mass.

\section{Gamma-ray constraints}

In additions to neutrinos, decaying/annihilating DM particles produce primary photons (for instance in the decays of neutral pions), electrons and positrons as well. Due to the propagation in the intergalactic medium, the primary photon spectrum is degraded in energy. On the other hand, the $e^\pm$ produce secondary photons through inverse Compton scatterings on the CMB. These processes would lead to a photon spectrum in the GeV--TeV energy range that can be potentially observed in gamma-ray Telescope, like Fermi-LAT. Hence, the {\it multi-messenger} analyses can provided significant constraints on the neutrino flux for both astrophysical sources and DM particles. In particular, as already stated in the Introduction, the astrophysical neutrino flux produced via $p$--$p$ interactions can have a spectral index as large as 2.2, otherwise the Fermi-LAT spectrum would be overpopulated~\cite{Murase:2013rfa,Bechtol:2015uqb,Chakraborty:2016mvc}. As for the case of astrophysical sources, the DM neutrino flux would be potentially constrained by gamma-ray data.

The Fermi-LAT measurements provide a bound on the {\it integrated} electromagnetic energy density, defined as~$\omega_\gamma$, coming from decaying/annihilating DM signals. By integrating the measured isotropic diffuse gamma-ray background (IGRB) from 100~MeV up to 820~GeV~\cite{Ackermann:2014usa}, one obtains the experimental value for the electromagnetic energy density $\omega_\gamma^{\rm exp} \simeq  \left(4.0\pm0.7\right)\times10^{-7}\,{\rm eV/cm^3}$.

The total electromagnetic energy density injected by decaying/annihilating DM particles comes from the contributions of prompt gamma-rays and $e^\pm$ fluxes
\begin{equation}
\omega_\gamma^{\rm DM} = \frac{4\pi}{c}\int^{E_{\rm max}}_{E_{\rm min}} \sum_{i=\text{gal,extragal}} \left[ E_\gamma \left.\frac{{\rm d}\phi_\gamma}{{\rm d}E_\gamma}\right|^i +  E_{e} \left.\frac{{\rm d}\phi_{e}}{{\rm d}E_{e^\pm}}\right|^i\right]dE \,,
\label{eq:em}
\end{equation}
where the galactic contribution is related to the anti-Galactic Center flux ($b=0$ and $\ell=\pi$ in Galactic coordinates). Such an integral is performed from $E_{\rm min} = 0.1$~GeV to $E_{\rm max}$ that corresponds to $m_{\rm DM}/2$ and $m_{\rm DM}$ in case of decaying and annihilating DM particles, respectively. The quantity $\omega_\gamma^{\rm DM}$ cannot overcome $\omega_\gamma^{\rm exp}$, independently of the propagation in the intergalactic medium, and this provides a robust constraint $\left(\omega_\gamma^{\rm DM} \leq \omega_\gamma^{\rm exp}\right)$ on DM viable models, hereafter defined as IGRB bound. This multi-messenger approach proposed in Ref.~\cite{Esmaili:2014rma} would lead to conservative constraints on the parameters (mass, lifetime, cross section) defining DM models (see also Ref.s~\cite{Murase:2012xs,Esmaili:2015xpa} for more dedicated multi-messenger approaches). Taking for example the bounds reported in Eqs.~(5.3)~and~(5.6) of Ref.~\cite{Murase:2012xs}, and updating the measurements of cosmological parameters and the IGRB spectrum, one recovers the same bounds shown in the plots of the next Section, once the same clumpiness factor is used.

\section{Method and Results}

In the present analysis, we try to statistically quantify the preference of the IC data for a two-components scenario, where one of them has a DM origin, with respect to a single astrophysical component. To this aim we adopt a likelihood-ratio test where the Test Statistics (TS) is given by
\begin{equation}
{\rm TS} = 2 \ln \frac{\mathcal{L}\left(\phi^{\rm Astro}_0,\phi^{\rm DM}_0\neq0\right)}{\mathcal{L}\left(\phi^{\rm Astro}_0,\phi^{\rm DM}_0=0\right)} \,,
\label{eq:TS}
\end{equation}
where $\phi^{\rm Astro}_0$ is defined in Eq.~\eqref{eq:astro}, and $\phi^{\rm DM}_0$ is the normalization of the DM neutrino flux, i.e. the inverse lifetime $1/\tau_{\rm DM}$ and the thermally averaged cross section $\left<\sigma v \right>$ in case of decaying and annihilating DM, respectively. The likelihood function $\mathcal{L}$ adopted in Eq.~\eqref{eq:TS} is a binned multi-Poisson likelihood~\cite{Baker:1983tu}, whose expression is equal to
\begin{equation}
\ln \mathcal{L} = \sum_i\left[n_i - N_i + n_i \ln\left(\frac{N_i}{n_i}\right)\right]\,,
\end{equation}
where the expected number of events $N_i$ is compared with the observed number of neutrinos $n_i$, once the background events (conventional atmospheric neutrinos and penetrating muons) have been subtracted in each bin $i$.

The number of events $N_i$ that is predicted by the total differential flux of Eq.~\eqref{eq:tot_flux} in the bin $i$ (defined by the deposited energy range $\Delta E'_i$ and by the reconstructed solid angle $\Delta\Omega'_i$) is obtained by the following integral
\begin{equation}
N_i = \Delta t \int_{\Delta E'_i} dE'_\nu \int_{\Delta \Omega'_i} d\Omega' \int dE_\nu \int_{4\pi} d\Omega \, \sum_\alpha \frac{{\rm d}J_\alpha}{{\rm d}E_\nu {\rm d}\Omega} \, A_\alpha \left(E_\nu,\Omega;E'_\nu,\Omega'\right) \,,
\label{eq:number}
\end{equation}
where $\Delta t = 641$~days is the exposure time of the 2-years MESE data sample and the quantity $A_\alpha \left(E_\nu,\Omega;E'_\nu,\Omega'\right)$ stands for the effective area of the IC detector per neutrino flavour $\alpha$\footnote{The IceCube effective area can be found at \url{https:/icecube.wisc.edu/science/data/HEnu_above1tev}.}. Such a quantity is a function of the neutrino energy $E_\nu$, the deposited one $E'_\nu$ as well as of the angular coordinates $\Omega$ and the reconstructed ones $\Omega'$. Note that the reconstructed angle provided by IceCube is divided into northern $\left(-1.0\leq\cos\theta_{\rm rec}\leq0.2\right)$ and southern $\left(0.2\leq\cos\theta_{\rm rec}\leq1.0\right)$ hemispheres, where $\theta$ is the Zenith angle. In the expression of Eq.~\eqref{eq:number}, the sum over neutrinos and anti-neutrinos is implicitly considered.

The Test Statistics of Eq.~\eqref{eq:TS} is evaluated by fixing the particular DM model (final states and DM mass $m_{\rm DM}$) and the spectral index $\gamma$ of the astrophysical contribution,  and by scanning over the only reaming free parameter $\phi^{\rm DM}_0$. The quantity $\phi^{\rm Astro}_0$ is always fixed at the best-fit for the two values of the spectral index considered $\left(\gamma=2.0,2.2\right)$. Such approach, described in a slightly different framework in Ref.s~\cite{Ackermann:2013uma,Aisati:2015vma}, signals how the fit of MESE data is improved by adding a DM neutrino flux to the simplest model of a single power-law.

Thanks to the Wilks theorem~\cite{Wilks:1938dza} and the Chernoff one~\cite{Chernoff}, the square root of TS represents the number of standard deviations $\sigma$ that the data prefer the two-components scenario with respect to the single astrophysical power-law. The larger the significance, more likely the presence of a second DM component. A similar analysis was performed for searches of DM gamma-lines in the Fermi-LAT spectrum~\cite{Ackermann:2013uma}, and very recently it has been proposed to characterize the significance of DM neutrino-lines in IceCube~\cite{Aisati:2015vma}.

In the following we will scrutinize different DM models (decay and annihilation), having different neutrino energy spectra (quarks or leptons in the final states) and different DM halo density distributions (NFW and ISO galactic distributions). Moreover, we also consider different astrophysical scenarios, that are related to the different values of spectral index. In particular, this statistical test is performed for each of the two values for the spectral index~$\gamma$ and for different DM masses in the range 1~TeV--10~PeV.

\subsection{Decaying DM case: $\chi \to f \overline{f}$ and $\chi \to \ell^-\ell^+\nu$}

\begin{figure}[t!]
\begin{center}
\includegraphics[width=0.42\textwidth]{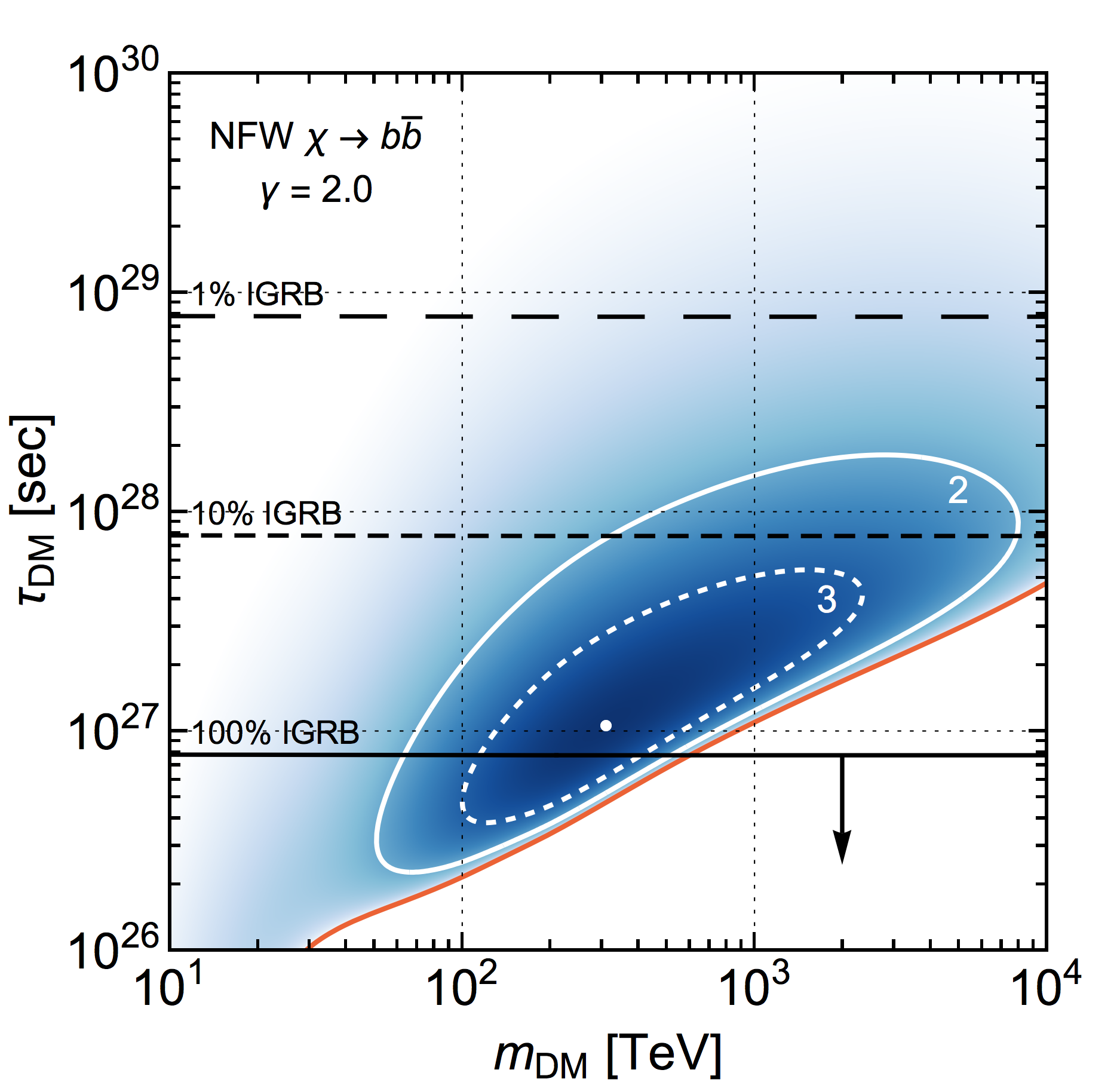}
\hskip2.mm
\includegraphics[width=0.42\textwidth]{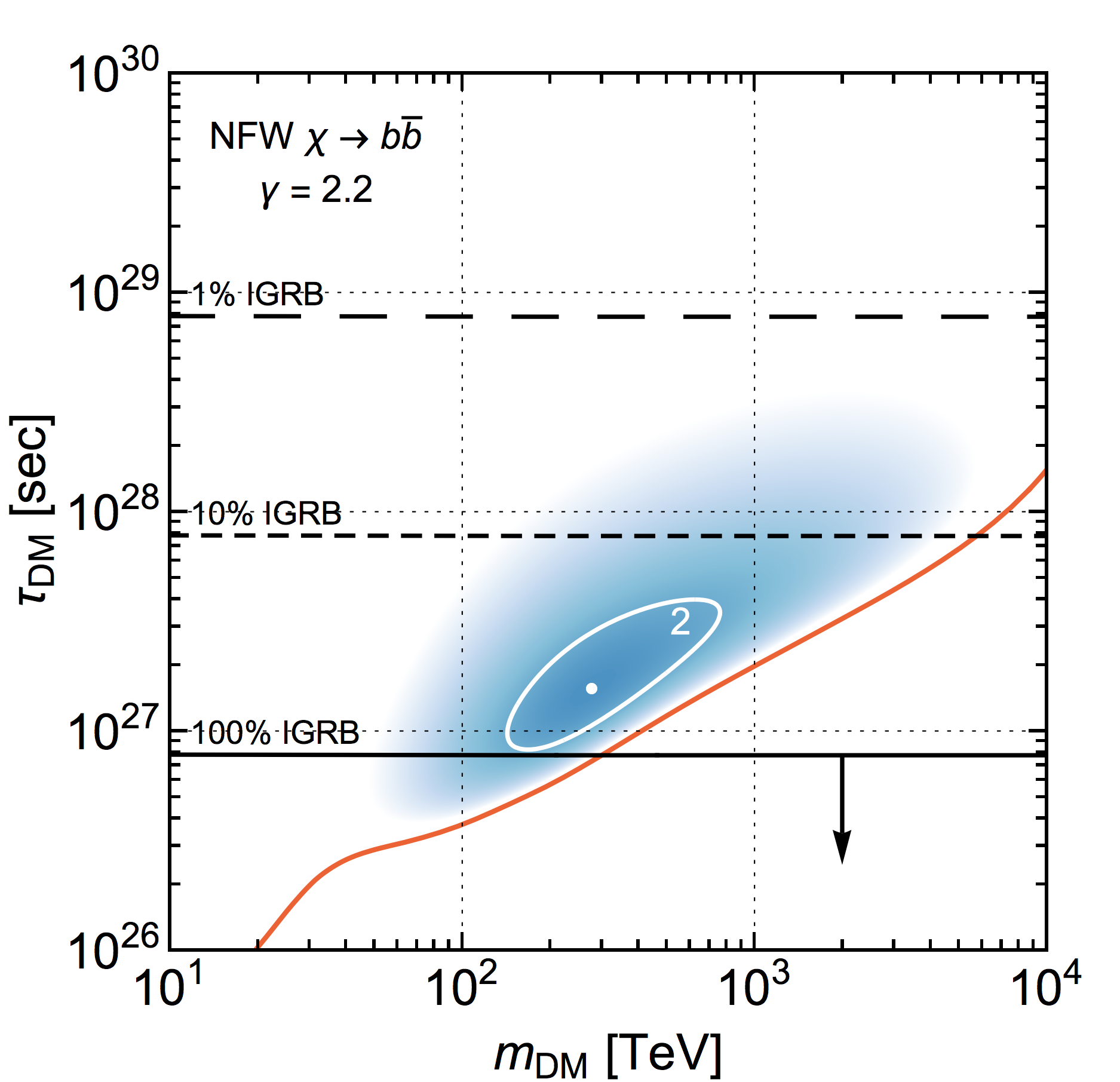}
\includegraphics[width=0.077\textwidth]{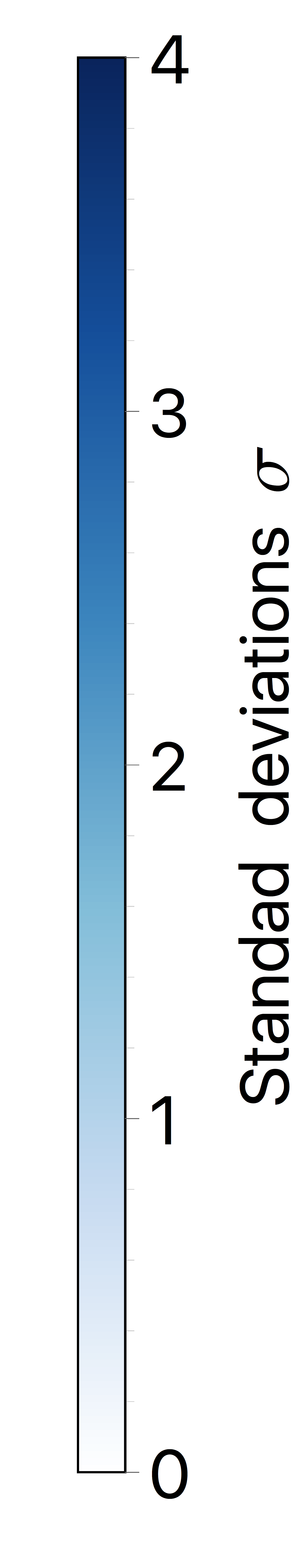}
\end{center}
\begin{center}
\includegraphics[width=0.42\textwidth]{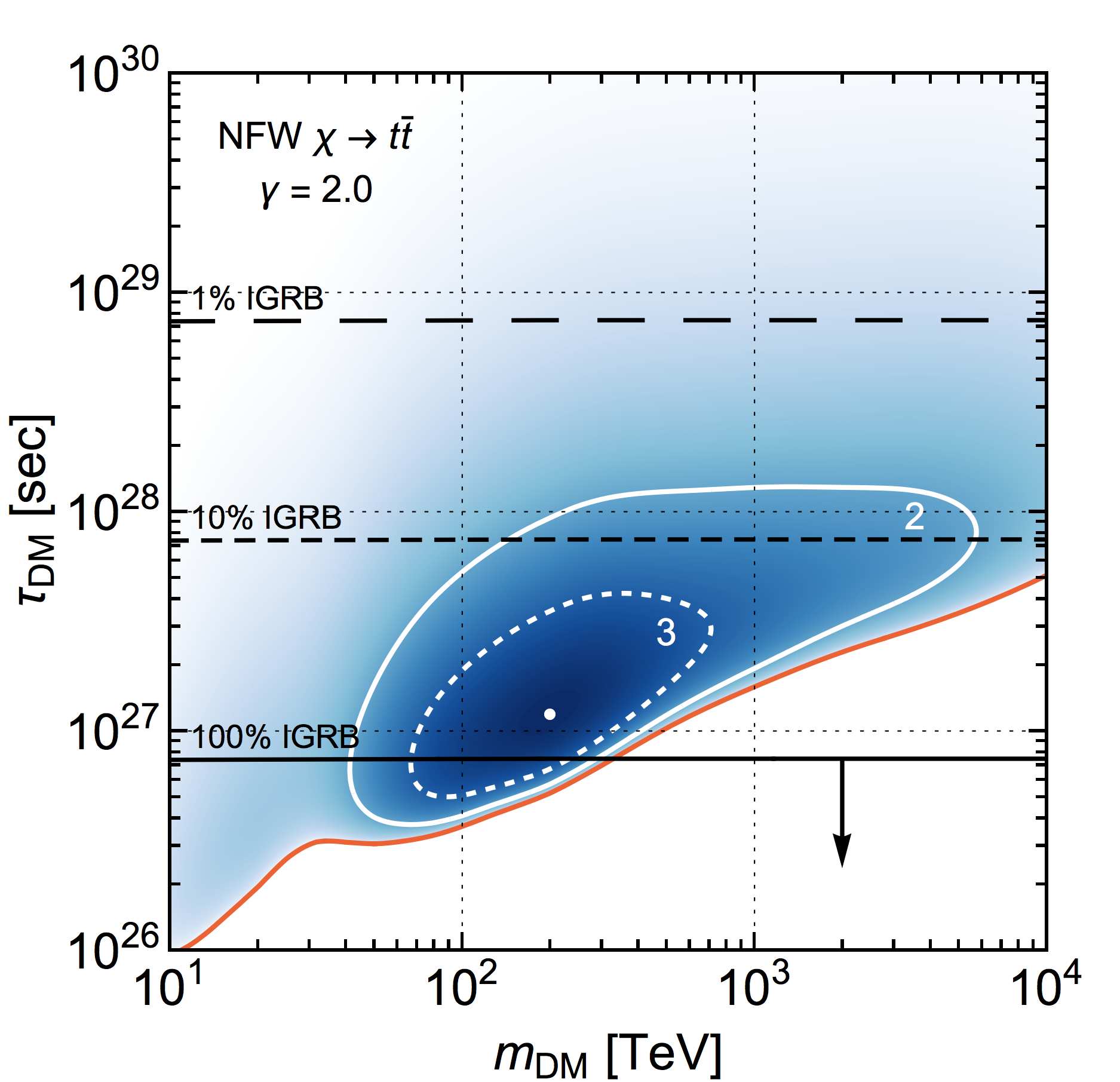}
\hskip2.mm
\includegraphics[width=0.42\textwidth]{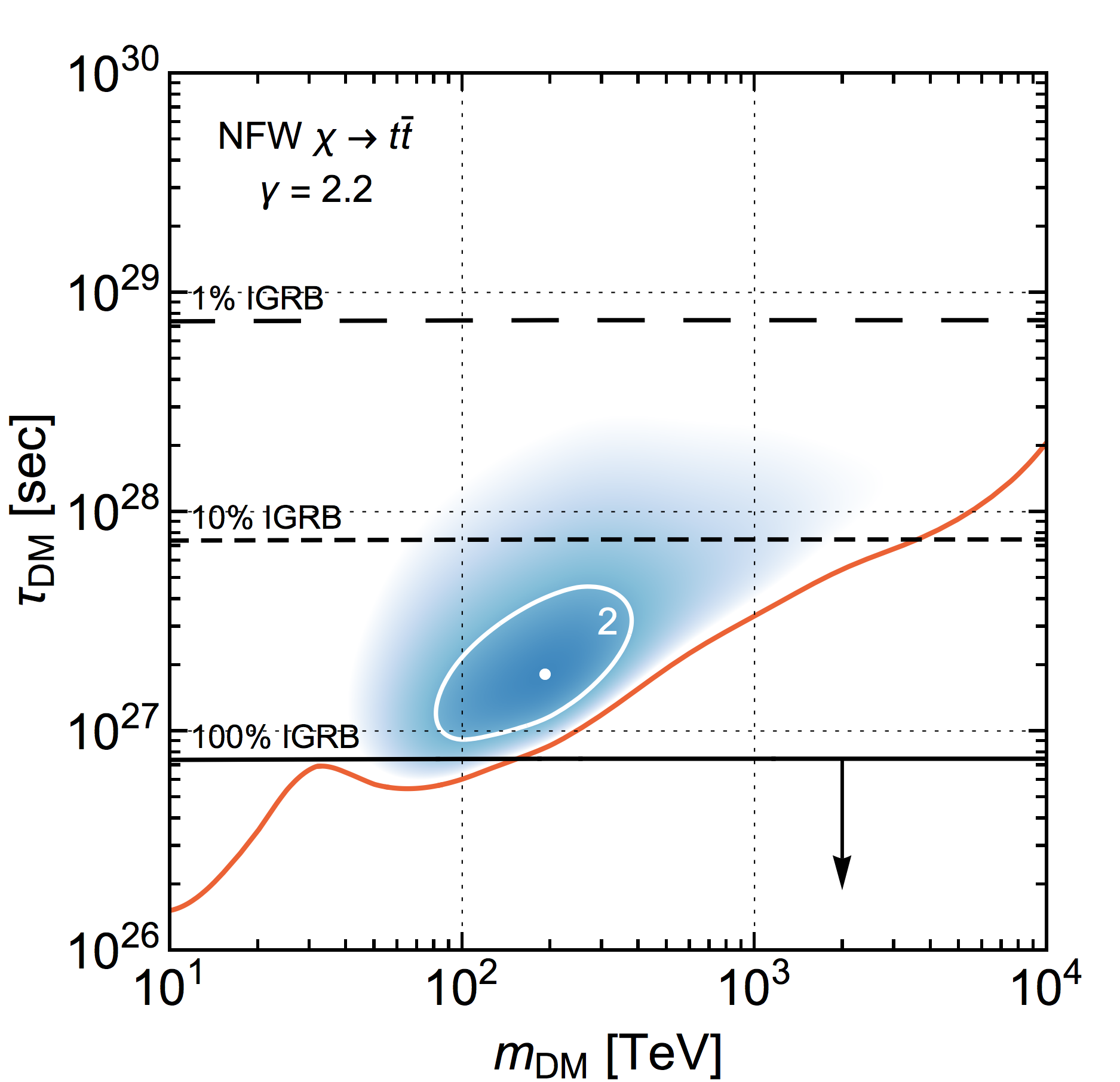}
\includegraphics[width=0.077\textwidth]{bar.png}
\end{center}
\caption{\label{fig:dec_quark}Number of standard deviations in $\sigma$ in the $m_{\rm DM}$--$\tau_{\rm DM}$ plane in case of decaying DM into SM quarks, $\chi \to b \overline{b}$ (upper panels) and $\chi \to t \overline{t}$ (lower panels), once the spectral index of the astrophysical power-law  has been fixed to 2.0 (left panels) and 2.2 (right panels). The white contours refer to $2\sigma$ (solid) and $3\sigma$ (dashed) significance level, and the white dot is the best-fit. The red line bounds from below the allowed region according to IceCube data, while the black one delimits from above the region excluded by Fermi-LAT data (see Section~3).}
\end{figure}

In the present analysis, we study the scenario where neutrinos are produced by scalar DM particles $\chi$ decaying into a couple of SM fermions $f=b,t,\mu,\tau$ (only one channel at a time is considered)\footnote{In this paper, we focus on DM masses larger than 10~TeV in order to explain the IC excess. Bounds on decaying DM scenarios for $m_{\rm DM} \leq 10$~TeV are provided in Ref.~\cite{Abbasi:2011eq}.}. In Fig.~\ref{fig:dec_quark}, it is reported the number of standard deviations $\sigma$ in the $m_{\rm DM}$--$\tau_{\rm DM}$ plane in case of DM particles decaying into quarks, i.e. bottom quarks (upper panels) and top ones (lower panels). In particular, the darker the color, the larger the significance in $\sigma$ of the DM neutrino component. The left and right panels of the figure refer to an astrophysical power-law with spectral index 2.0 and 2.2, respectively. In the plots, the white dot is related to the maximum significance of the DM component, whereas the white contours corresponds to 2.0$\sigma$ (solid line) and 3.0$\sigma$ (dashed line), if present. Moreover, the red line bounds from above the region in the parameter space where the inclusion of a DM component to the neutrino flux makes the fit worse with respect to the case of a single astrophysical power-law. Therefore, the region delimited from such a red line is excluded by the present IC data, once a spectral index 2.0 and 2.2 is respectively assumed. Furthermore, the almost horizontal black lines show the gamma-ray constraints $\omega^{\rm DM}_\gamma \leq\omega^{\rm exp}_\gamma$ (see Section~3) related to different DM contributions (1\%, 10\% and 100\%) to the Fermi-LAT IGRB spectrum. In particular, the allowed region in the $m_{\rm DM}$--$\tau_{\rm DM}$ plane is bounded from below by the solid black line corresponding to $\omega^{\rm DM}_\gamma = \omega^{\rm exp}_\gamma$ (100\% IGRB). Such multi-messenger constraints are affected by an uncertainty of about 18\% according to the experimental value $\omega^{\rm exp}_\gamma$.

\begin{figure}[t!]
\begin{center}
\includegraphics[width=0.42\textwidth]{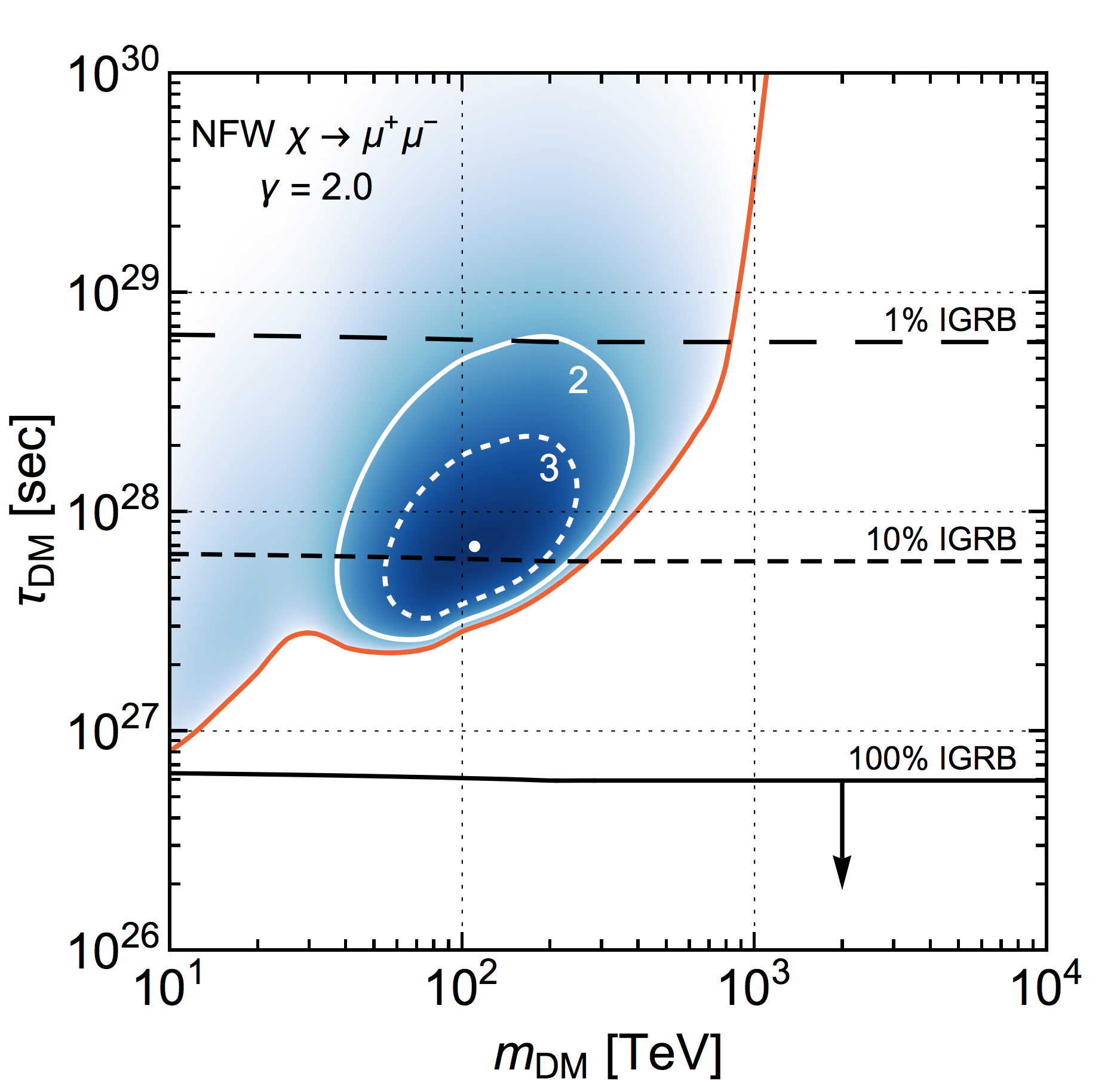}
\hskip2.mm
\includegraphics[width=0.42\textwidth]{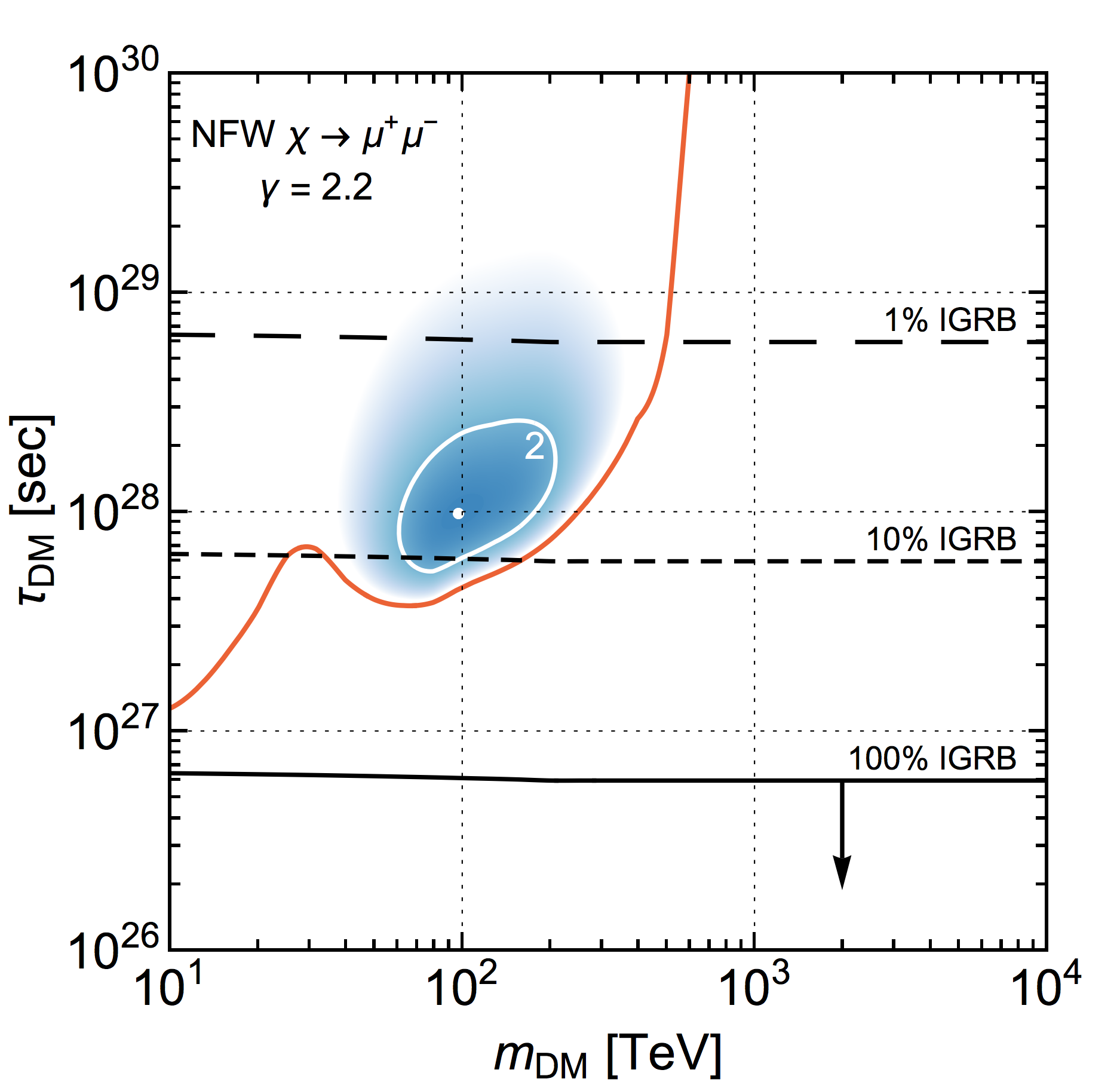}
\includegraphics[width=0.077\textwidth]{bar.png}
\end{center}
\begin{center}
\includegraphics[width=0.42\textwidth]{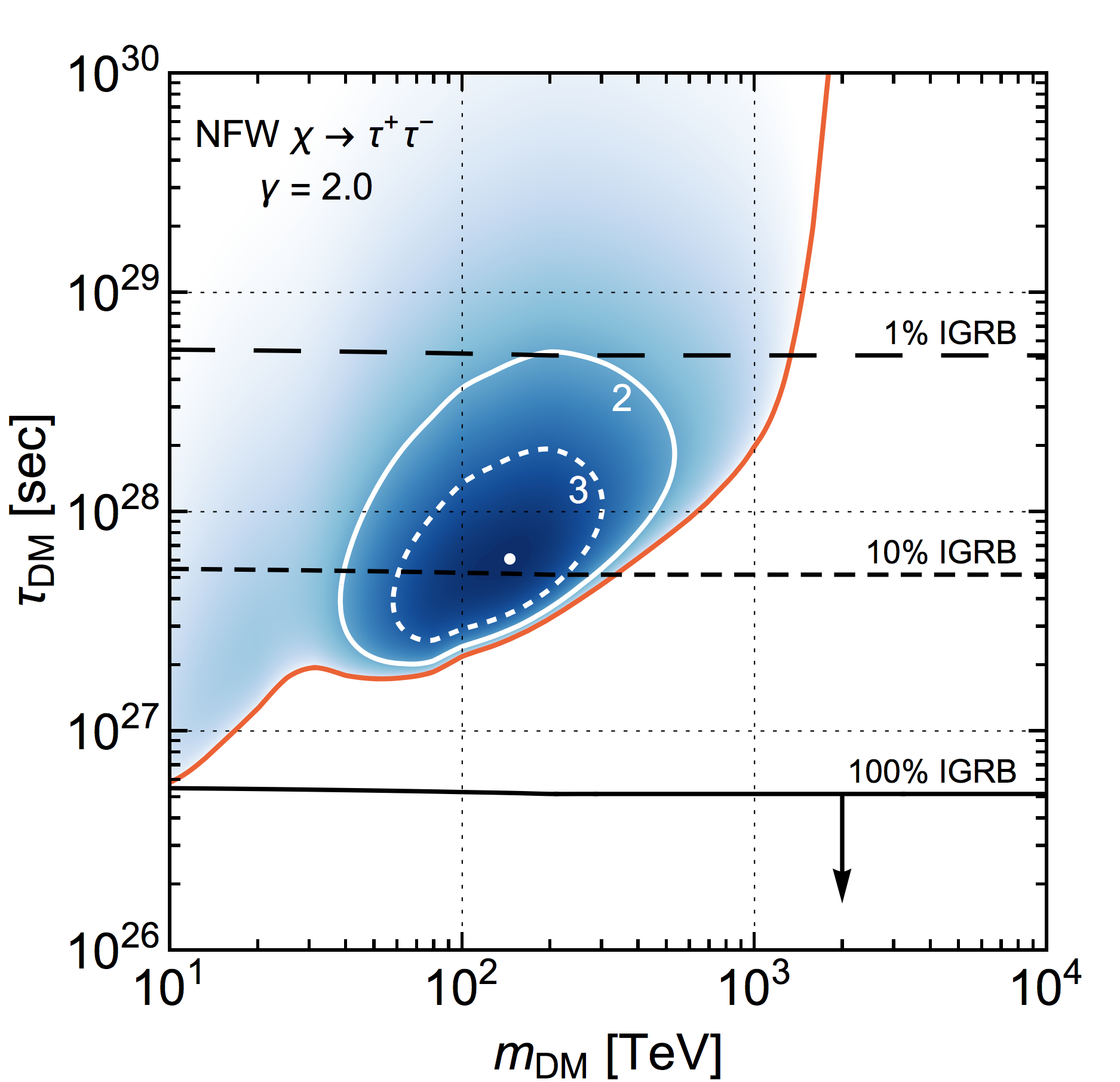}
\hskip2.mm
\includegraphics[width=0.42\textwidth]{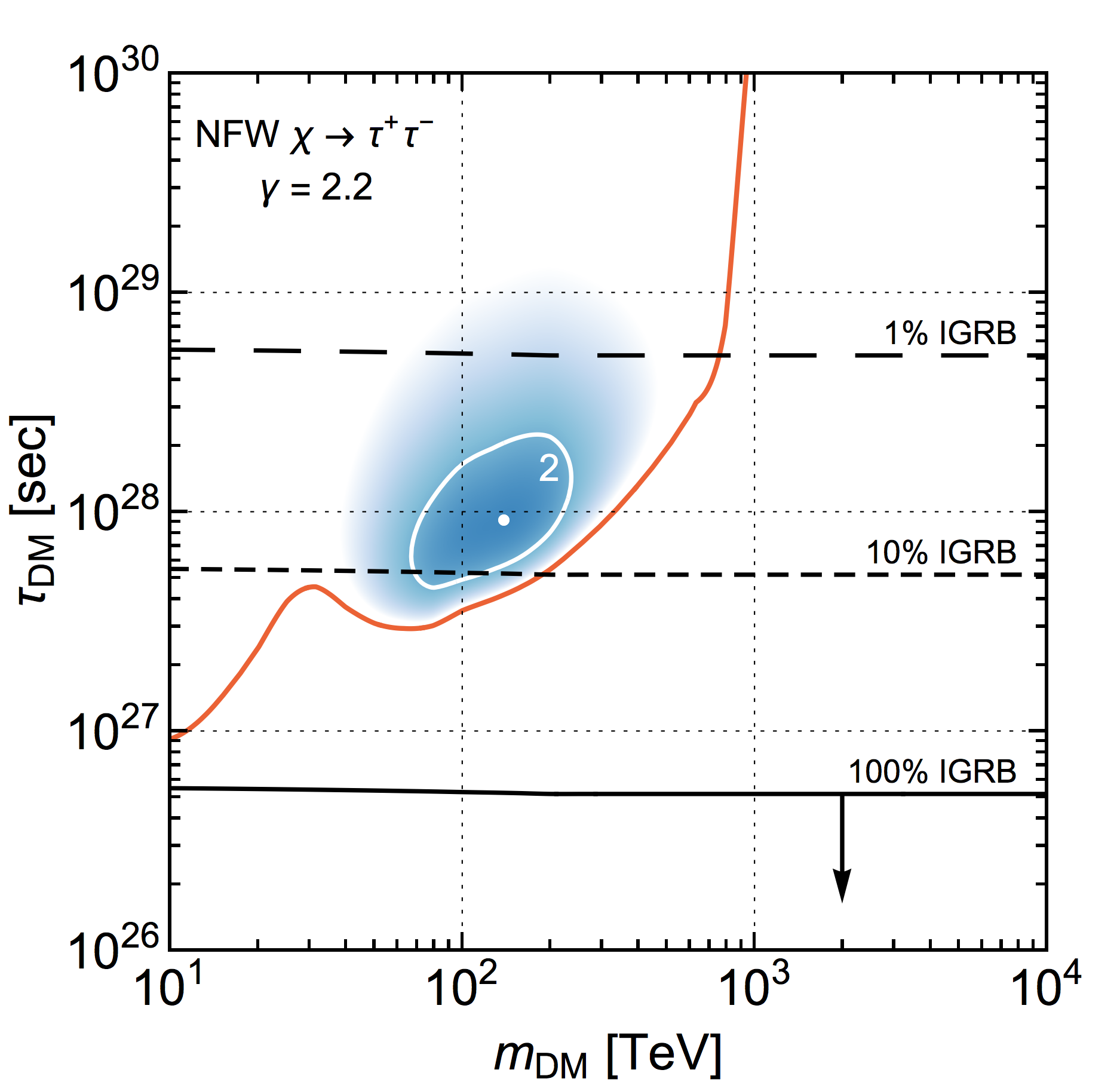}
\includegraphics[width=0.077\textwidth]{bar.png}
\end{center}
\caption{\label{fig:dec_lepton}Number of standard deviations in $\sigma$ in the $m_{\rm DM}$--$\tau_{\rm DM}$ plane in case of decaying DM into SM leptons, $\chi \to \mu^+\mu^-$ (upper panels) and $\chi \to \tau^+\tau^-$ (lower panels). The description of the plots is the same of Fig.~\ref{fig:dec_quark}.}
\end{figure}
\begin{figure}[t!]
\begin{center}
\includegraphics[width=0.42\textwidth]{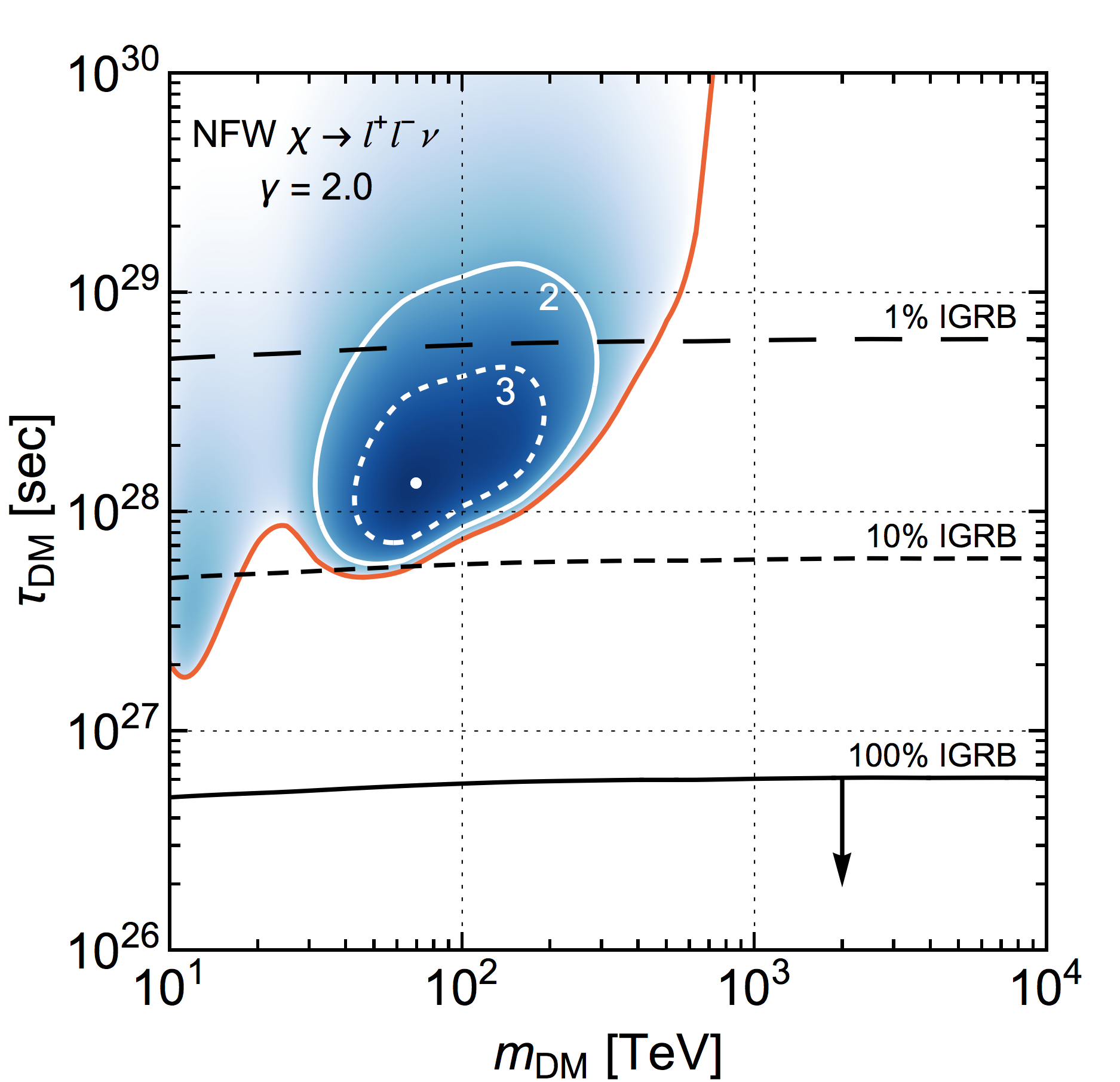}
\hskip2.mm
\includegraphics[width=0.42\textwidth]{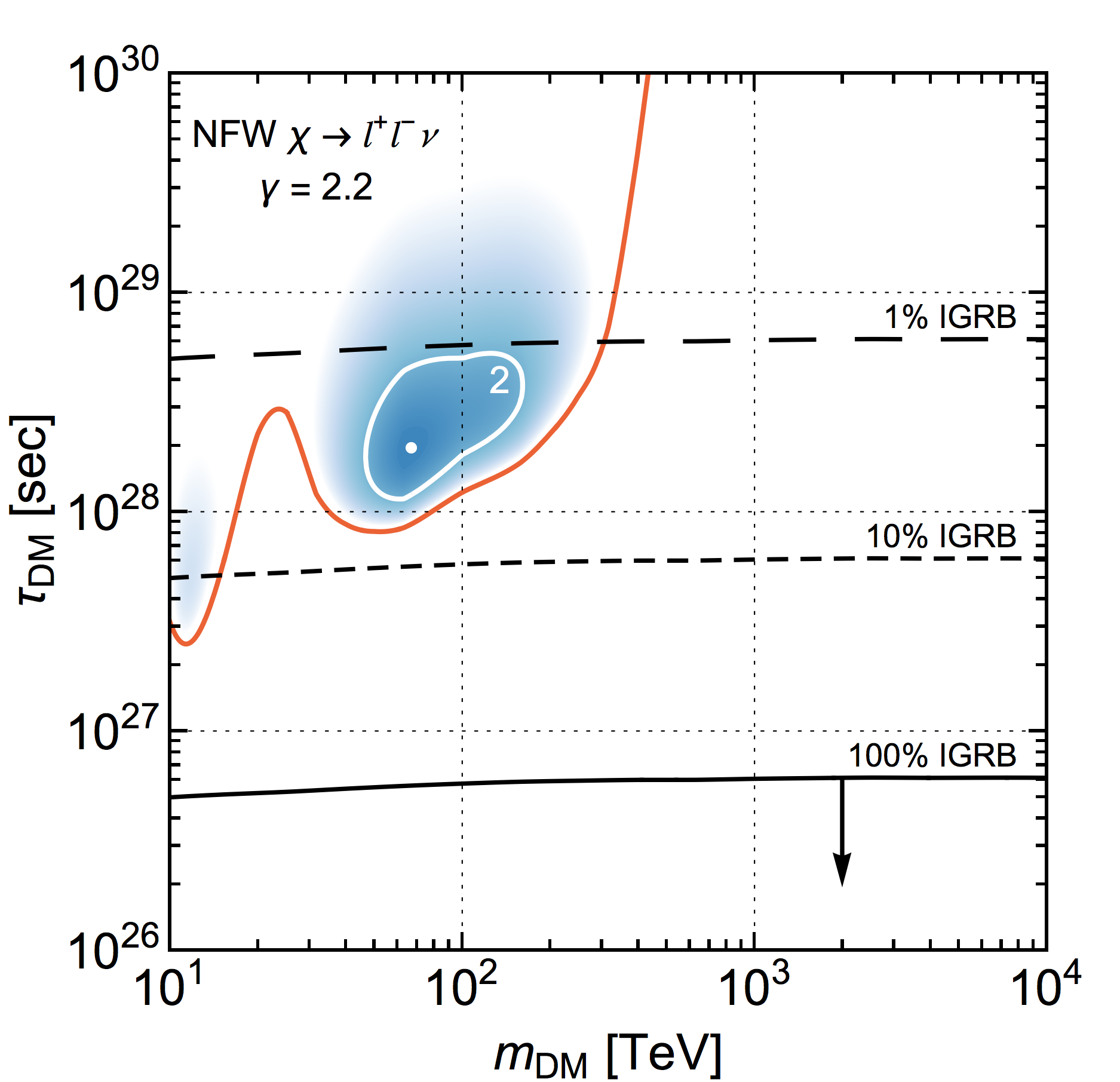}
\includegraphics[width=0.077\textwidth]{bar.png}
\end{center}
\caption{\label{fig:dec_A4}Number of standard deviations in $\sigma$ in the $m_{\rm DM}$--$\tau_{\rm DM}$ plane in case of DM model producing neutrinos via leptophilic three-bodies decays, $\chi \to \ell^+\ell^-\nu$. The description of the plots is the same of Fig.~\ref{fig:dec_quark}.}
\end{figure}

In Fig.~\ref{fig:dec_lepton}, the decaying cases into muon leptons (upper panels) and tau ones (lower panels) are reported. The case of leptophilic three-bodies decays ($\chi \to \ell^+\ell^-\nu$) is instead shown in Fig.~\ref{fig:dec_A4}. All the plots for the decaying DM cases are obtained by considering the Navarro-Frenk-White halo density distribution. The results related to the Isothermal distribution do not significantly differ from the NFW ones, and therefore they are not reported here. In particular, in case of ISO profile the statistical significance differs less than 1\% with respect to the NFW distribution. We observe that, in case of quarks in the final states (Fig.~\ref{fig:dec_quark}), smaller values for the lifetime $\tau_{\rm DM}$ and larger DM masses $m_{\rm DM}$ are favoured with respect to the case of leptonic final states (Fig.s~\ref{fig:dec_lepton}~and~\ref{fig:dec_A4}). Moreover, the models with quarks as final states are more in tension with the Fermi-LAT data with respect to the models involving leptons. In particular, in case of quark decay channels IceCube data prefer values of $m_{\rm DM}$ and $\tau_{\rm DM}$ close to the 100\% IGRB bound. This corresponds to the unrealistic situation where Fermi-LAT gamma-rays are completely explained in terms of a DM signal and not of astrophysical sources. On the other hand, in the case of a leptophilic DM, the most significant region in the parameter space $m_{\rm DM}$--$\tau_{\rm DM}$ corresponds to a IGRB contribution smaller than 10\%, situation implying that the Fermi-LAT gamma-ray observations are dominated by the astrophysical sources. Therefore, we can already conclude that in general the leptophilic scenarios are in fair agreement with both neutrinos and gamma-ray observations under the assumption of a two-components flux.

\begin{figure}[t!]
\centering
\includegraphics[width=0.45\textwidth]{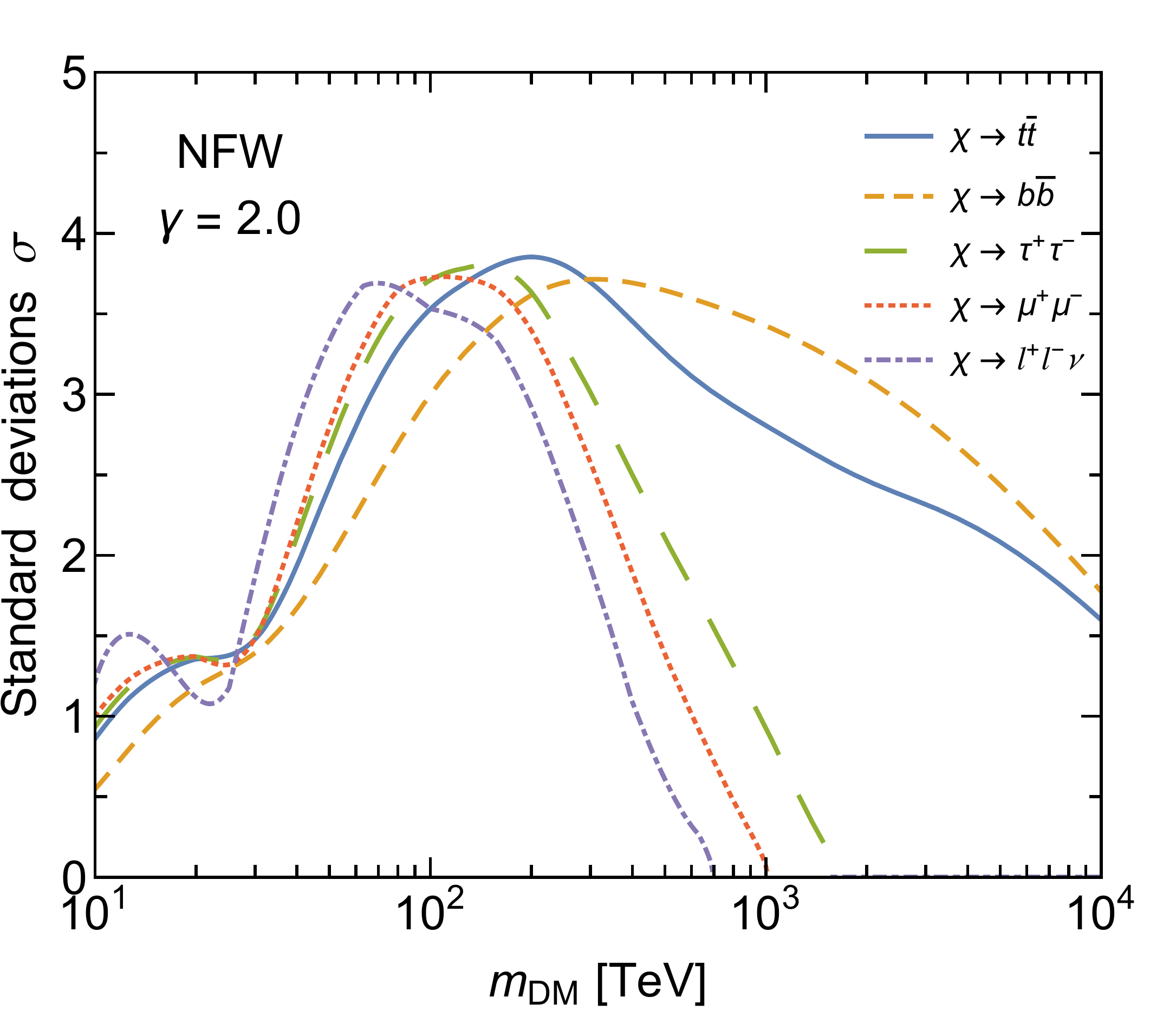}
\hskip3.mm
\includegraphics[width=0.45\textwidth]{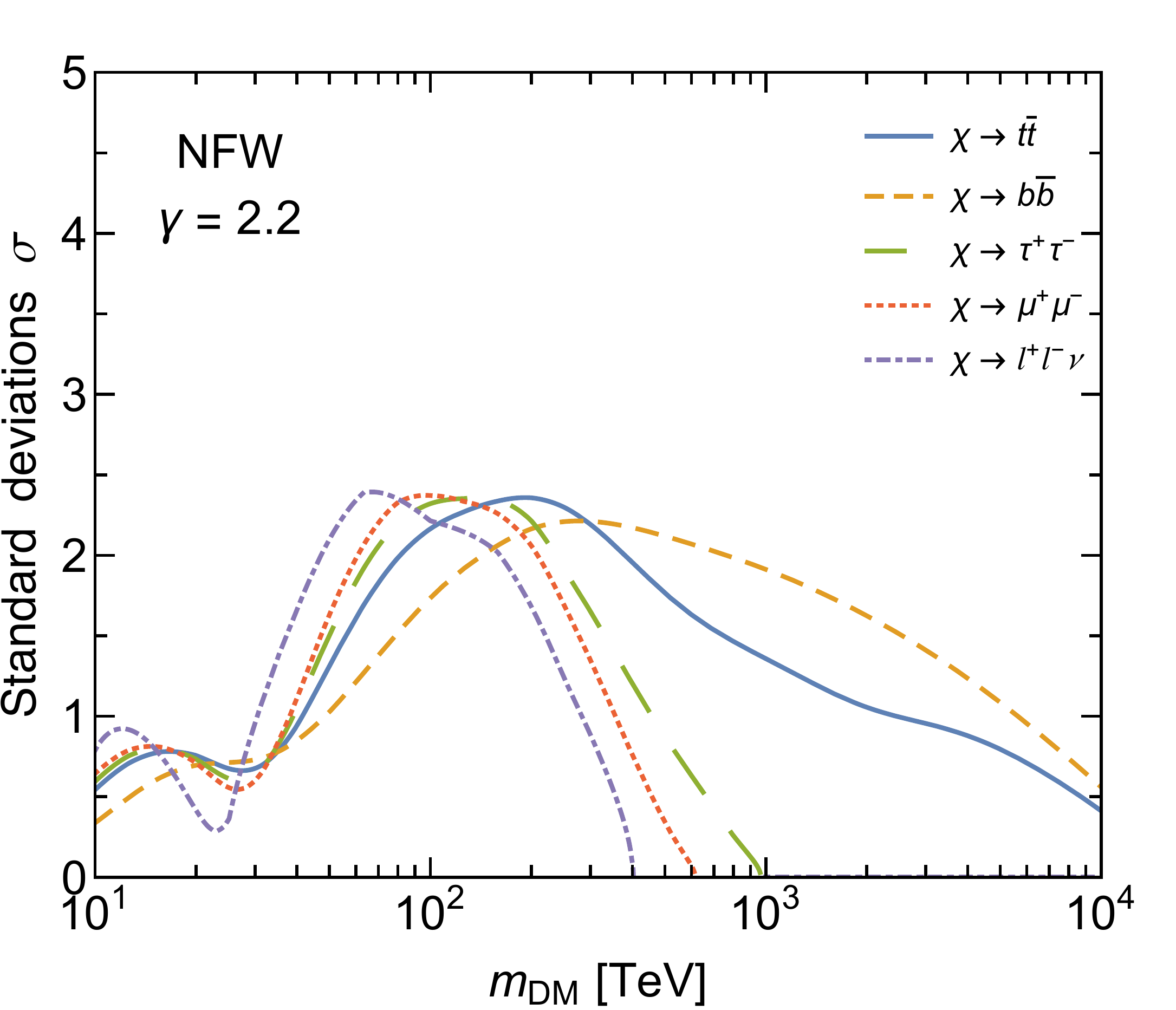}
\caption{\label{fig:dec}Significance in number of standard deviations $\sigma$ as a function of the DM mass $m_{\rm DM}$ for all the studied models of decaying DM, in case of a spectral index 2.0 (left panel) and 2.2 (right panel).}
\end{figure}

The significance in $\sigma$ as a function of DM mass $m_{\rm DM}$ is explicitly depicted in Fig.~\ref{fig:dec} for all the studied decaying cases. The curves shown in the plots have been obtained by considering the best-fit value of the DM lifetime for each DM model and each DM mass. As it is clear from the plots, the maximum value of $\sqrt{\rm TS}$ is almost independent on the decay channel considered and it results to be 3.7--3.9$\sigma$ and 2.2--2.4$\sigma$ in case of spectral index 2.0 and 2.2, respectively. Moreover, it is worth observing that the maximum significance is reached for $m_{\rm DM} \simeq 140$~TeV for a DM decaying mainly in leptons, while it is maximized at $m_{\rm DM} \simeq 200$~TeV and $m_{\rm DM} \simeq 300$~TeV for the cases $\chi \to t\overline{t}$ and $\chi \to b\overline{b}$, respectively. This is because neutrinos are mainly produced at low energy in the hadronic cascades, while in the leptonic channels their energy can be as large as $m_{\rm DM}/4$. This consideration also explains why DM masses larger than about 1~PeV (700~GeV) are excluded by IC data for the leptonic decay channels for $\gamma=2.0$ ($\gamma=2.2$), while no constraints are found in case of hadronic channels. Moreover, the smallest DM mass for the best-fit is obtained in case of the leptophilic three-bodies decays (dotdashed purple line in Fig.~\ref{fig:dec}). This is due to fact that in such a case primary neutrinos are produced up to an energy of $m_{\rm DM}/2$. 

\subsection{Annihilating DM case: $\chi \chi \to f\overline{f}$}

\begin{figure}[t!]
\begin{center}
\includegraphics[width=0.42\textwidth]{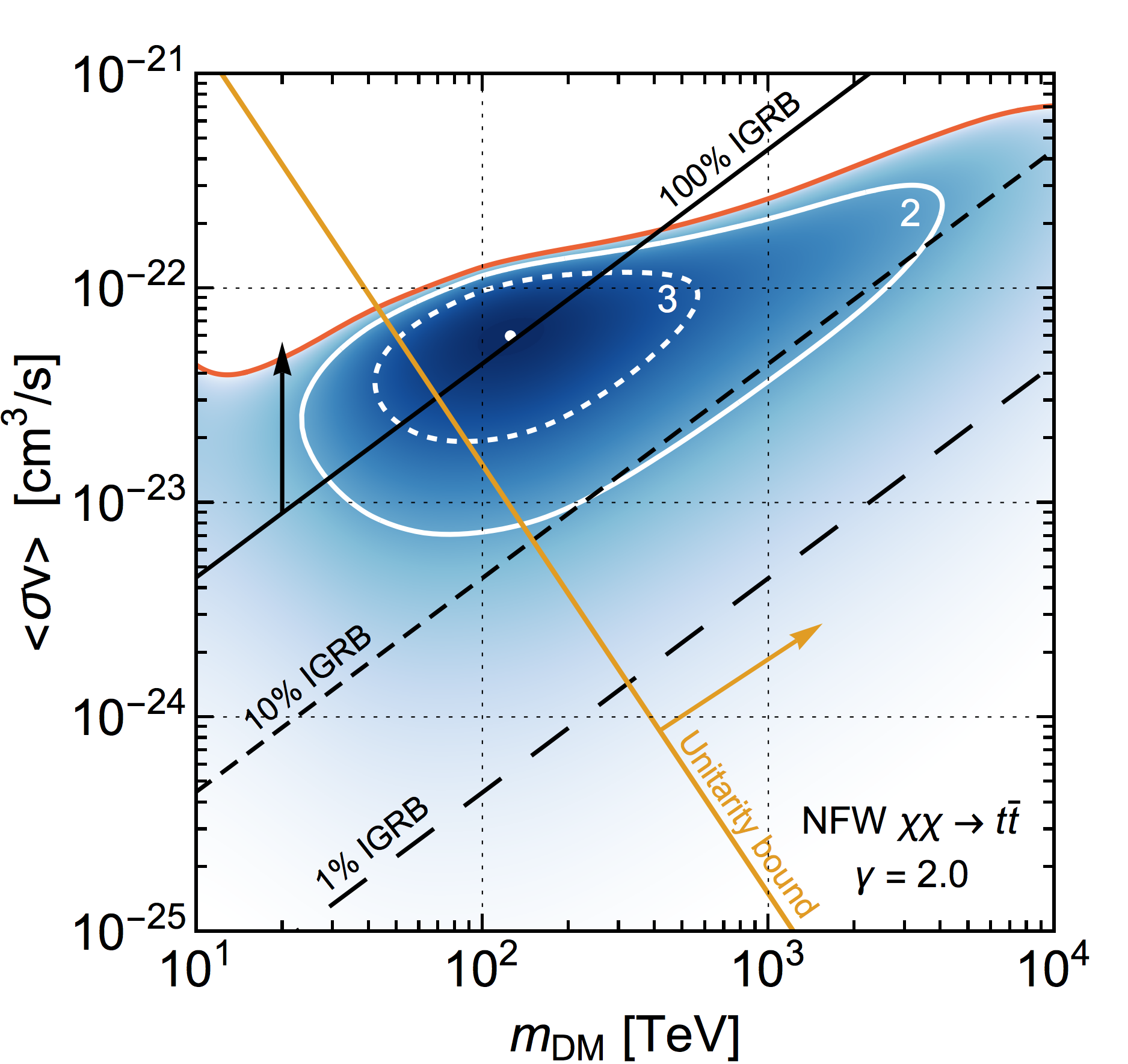}
\hskip2.mm
\includegraphics[width=0.42\textwidth]{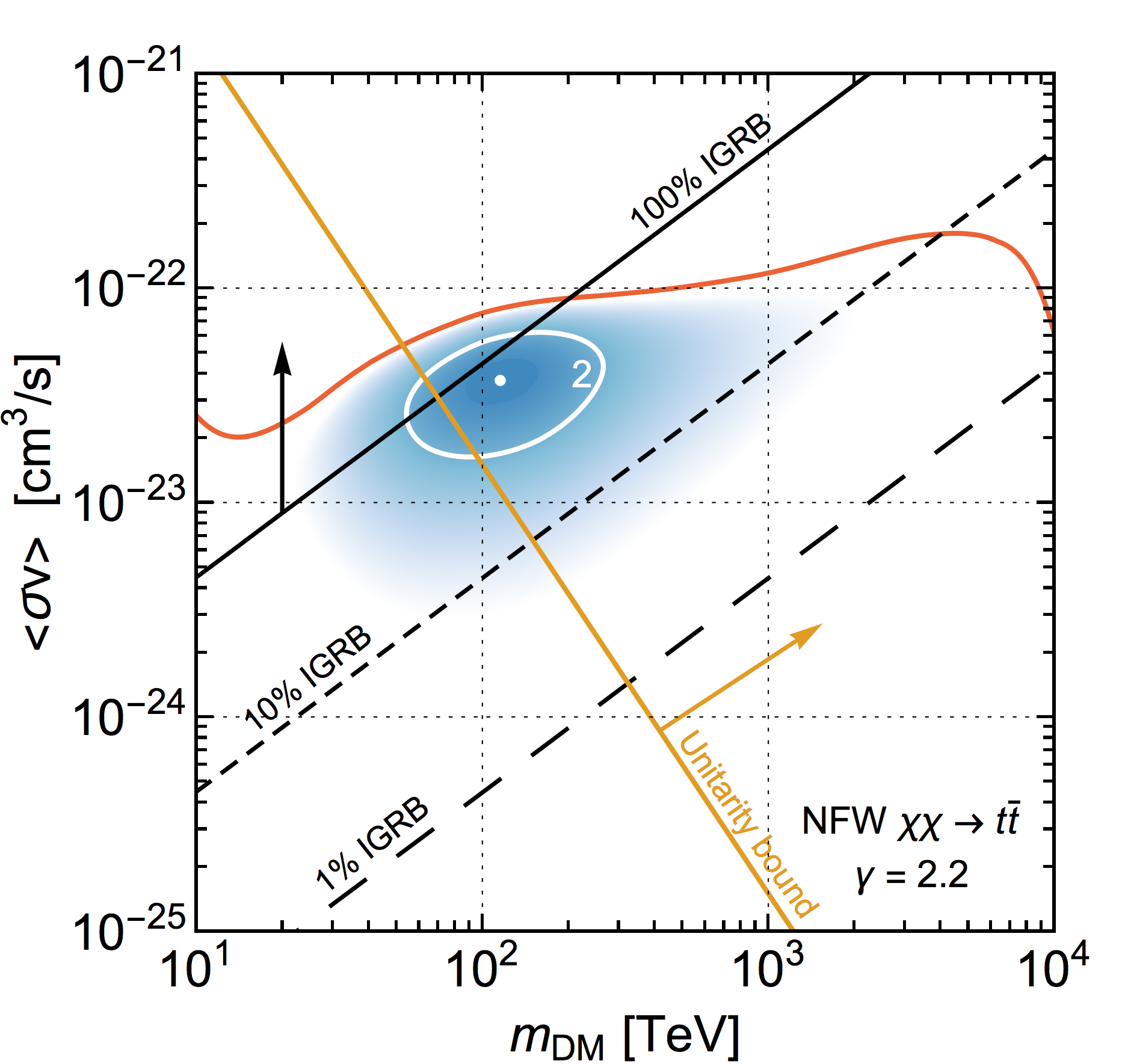}
\includegraphics[width=0.075\textwidth]{bar.png}
\end{center}
\begin{center}
\includegraphics[width=0.42\textwidth]{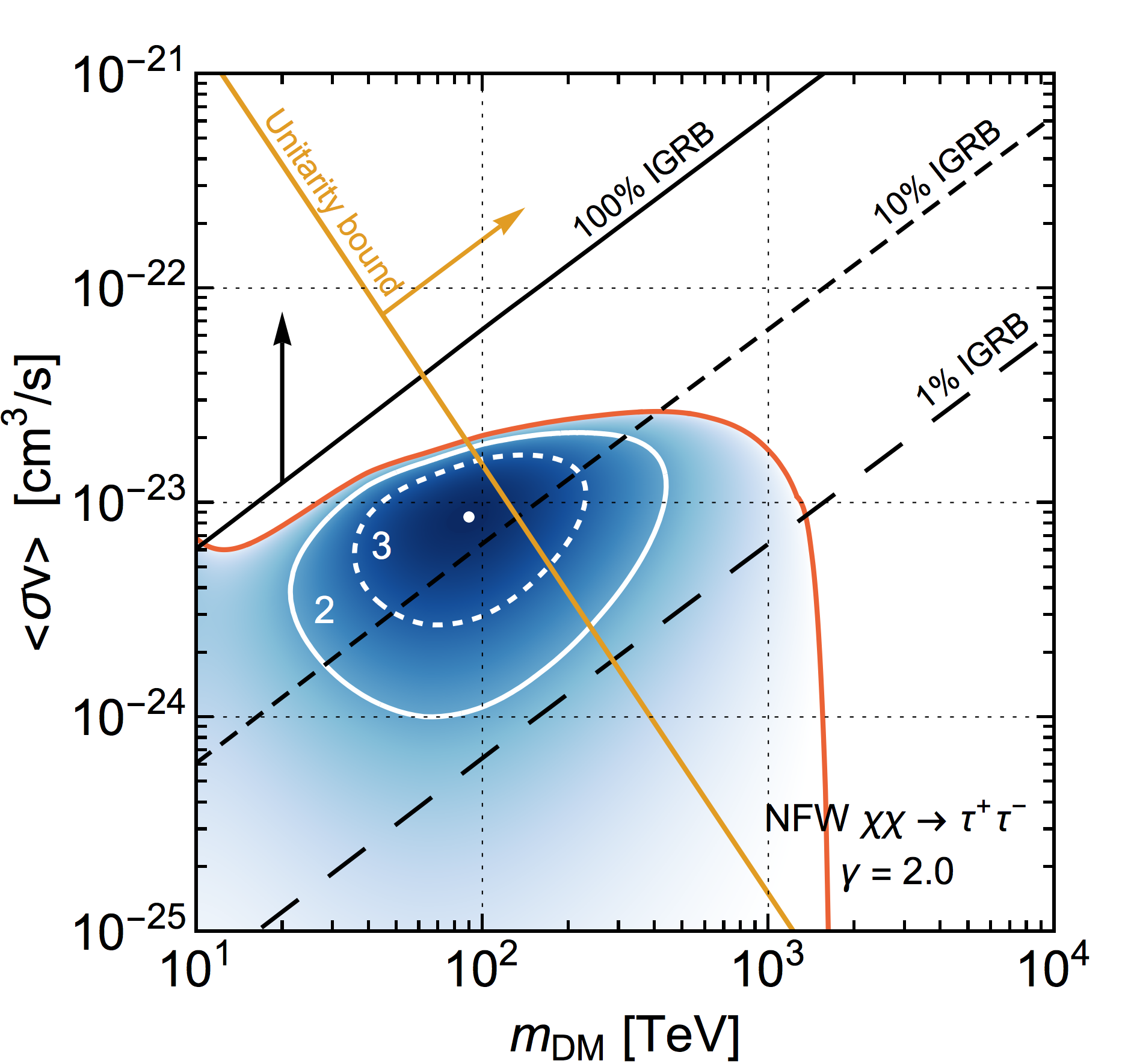}
\hskip2.mm
\includegraphics[width=0.42\textwidth]{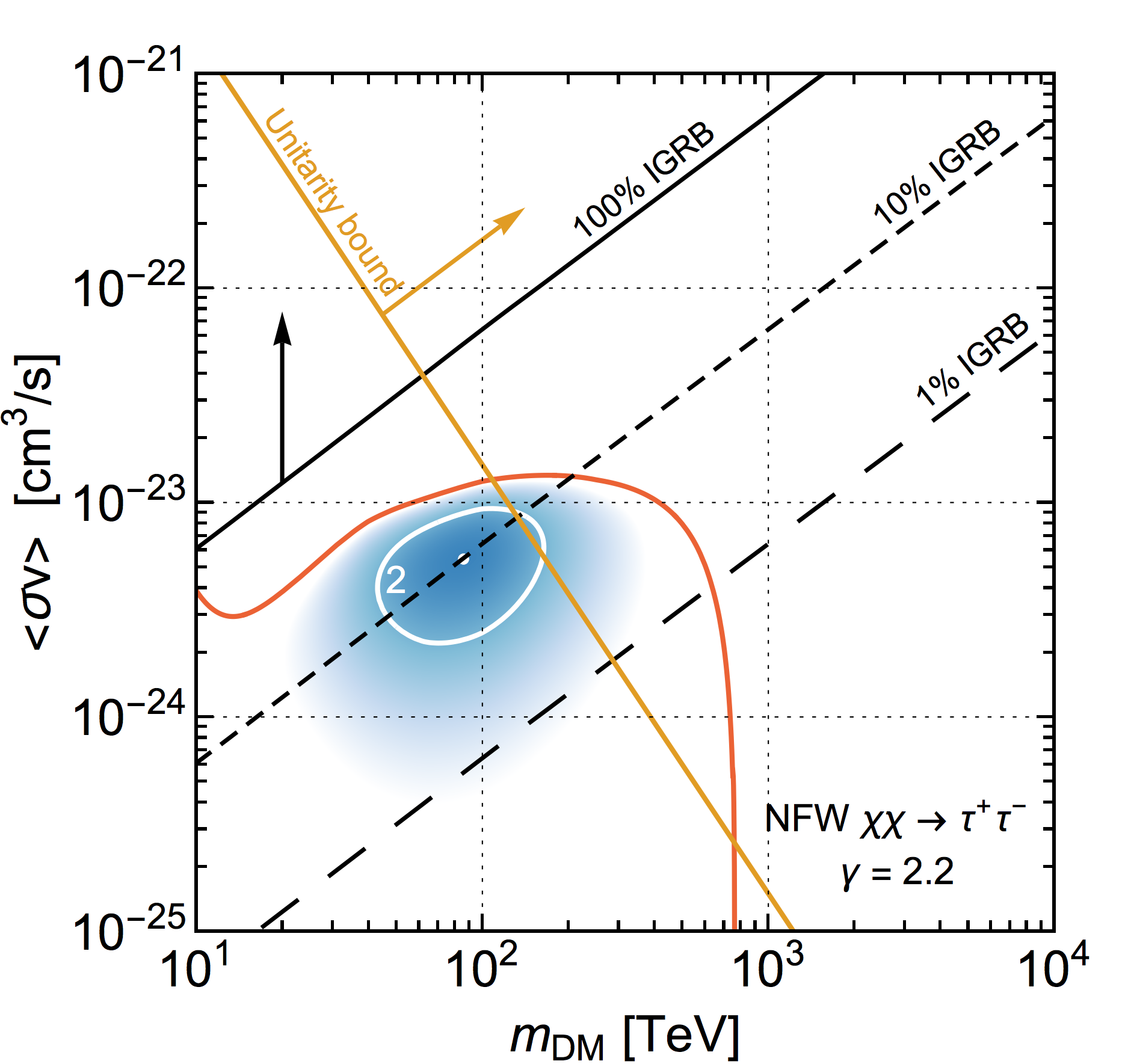}
\includegraphics[width=0.075\textwidth]{bar.png}
\end{center}
\caption{\label{fig:ann}Number of standard deviations in $\sigma$ in the $m_{\rm DM}$--$\left<\sigma v\right>$ plane in case of annihilating DM into top quarks $\chi \chi \to t\overline{t}$ (upper panels) and into tau leptons $\chi \chi \to \tau^+\tau^-$ (lower panels). The white contours refer to $2\sigma$ (solid) and $3\sigma$ (dashed) significance level, and the white dot represents the best-fit. The red line bounds from above the allowed region according to IceCube data, while the black solid one delimits from below the region excluded by Fermi-LAT data (see Section~3). The yellow line represents the unitarity constraint of the thermally averaged cross section provided in Eq.~\eqref{eq:unitarity}.}
\end{figure}

The results for the annihilating case\footnote{The bounds for the annihilating DM scenarios with $m_{\rm DM}\leq10$~TeV are provided in Refs.~\cite{Adrian-Martinez:2015wey,Aartsen:2016pfc}.} into top quarks ($\chi \chi \to t\overline{t}$) and tau leptons ($\chi \chi \to \tau^+\tau^-$) are depicted in Fig.~\ref{fig:ann}. Such two cases are representative for all the hadronic and leptonic scenarios. Hence, we do not report the cases $\chi \chi \to b\overline{b}$ and $\chi \chi \to \mu^+\mu^-$. The best-fit points (white dots) are reached for the same values of the DM mass obtained in the decaying case, and for a thermally averaged cross section of the order of $10^{-24}$--$10^{-23}$~${\rm cm^3/s}$. As for the decaying case, the maximum significance in standard deviations is $\sim3.8\sigma$ and $\sim 2.3\sigma$ for spectral index 2.0 and 2.2, respectively. The significance level of the two-components flux decreases by 0.1 if the ISO distribution rather than the NFW one is considered. A much larger dependence on the DM halo profile would be expected in case of a spatial study on neutrino events that would require a larger statistics and a detailed knowledge of the IC effective area.

In addition to the gamma-rays constraints (black lines), that bound from above the allowed region of the parameter space, too large values for $\left<\sigma v\right>$ are excluded according to the {\it unitarity bound}~\cite{Griest:1989wd,Hui:2001wy,Kusenko:2001vu} (yellow line). In particular, the thermally averaged cross section has to be smaller than
\begin{equation}
\left<\sigma v\right> \leq \frac{4\pi}{m_{\rm DM}~v} = 1.5 \times 10^{-23} \frac{\rm cm^3}{\rm s} \left[ \frac{\rm 100 \, TeV}{m_{\rm DM}}\right]^2\,,
\label{eq:unitarity}
\end{equation}
where the typical value of the DM velocity $v$ in the halo has been considered ($v = 300~{\rm km/s}$). As shown in the plots, DM particles annihilating into quarks are excluded by unitarity, since a detectable signal in IceCube requires larger values for $\left<\sigma v\right>$. On the other hand, leptophilic scenarios are compatible both with the IGRB constraints and with unitarity.

\section{Conclusions}

\begin{figure}[t!]
\begin{center}
\includegraphics[width=0.42\textwidth]{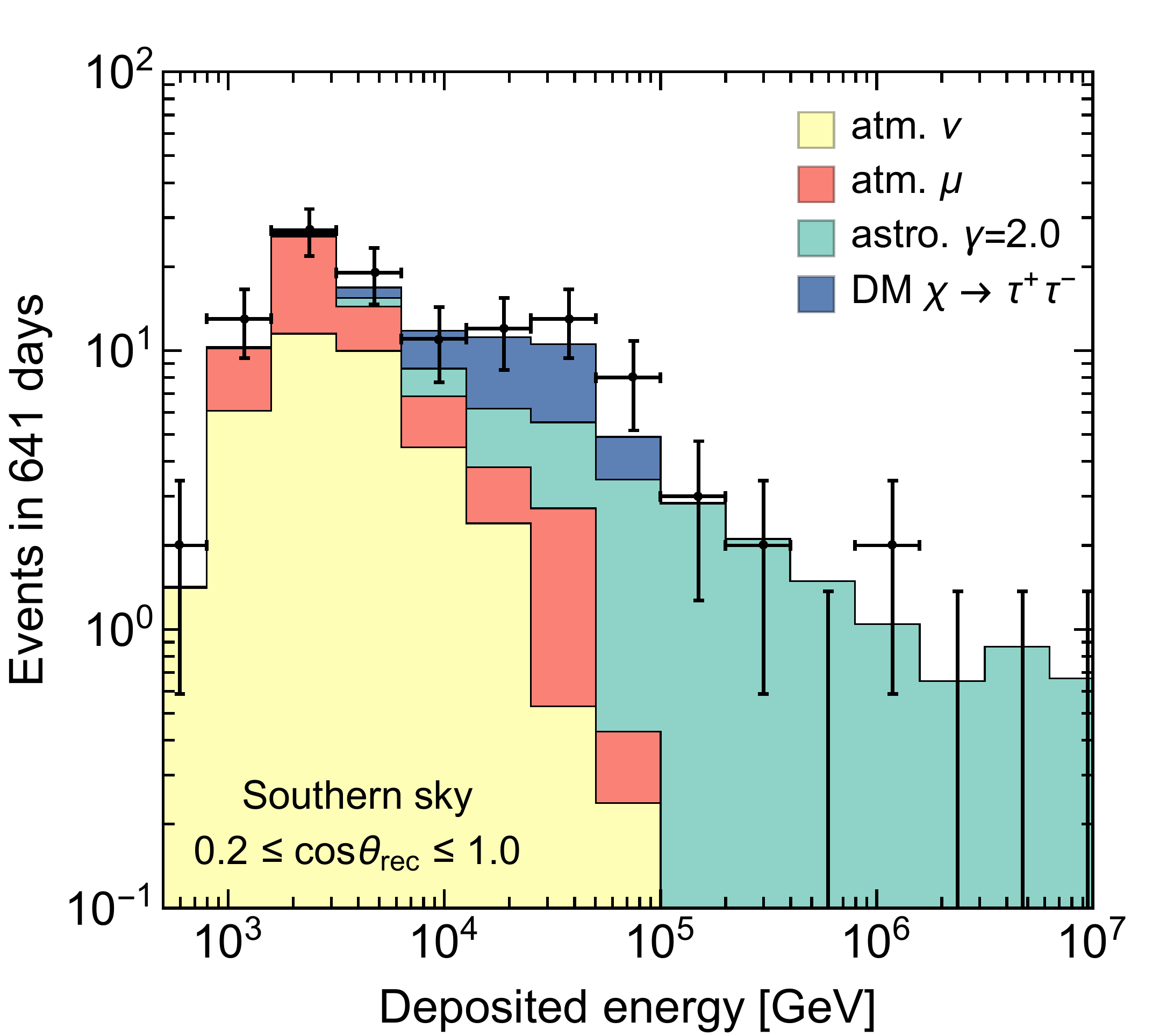}
\hskip2.mm
\includegraphics[width=0.42\textwidth]{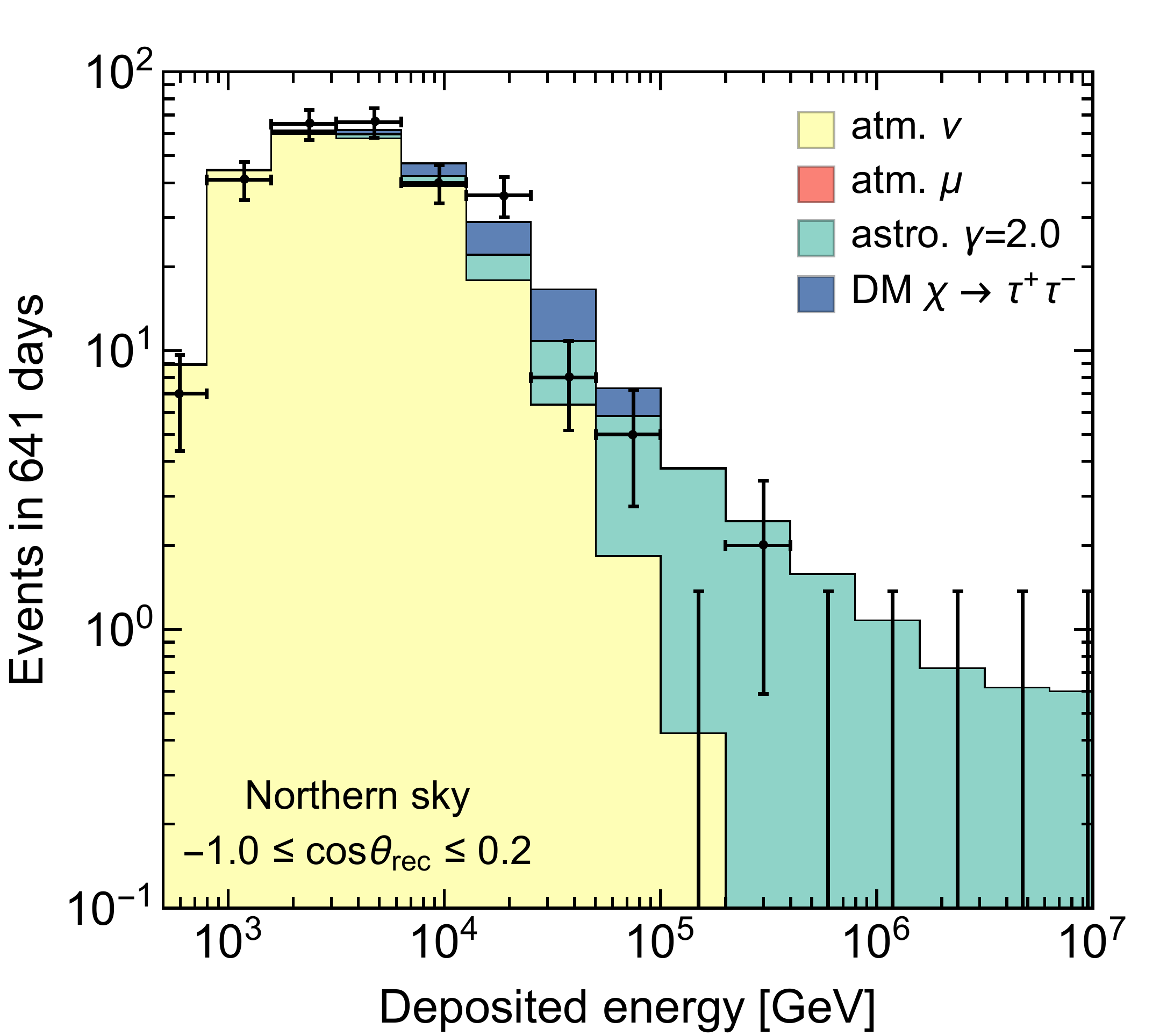}
\end{center}
\begin{center}
\includegraphics[width=0.42\textwidth]{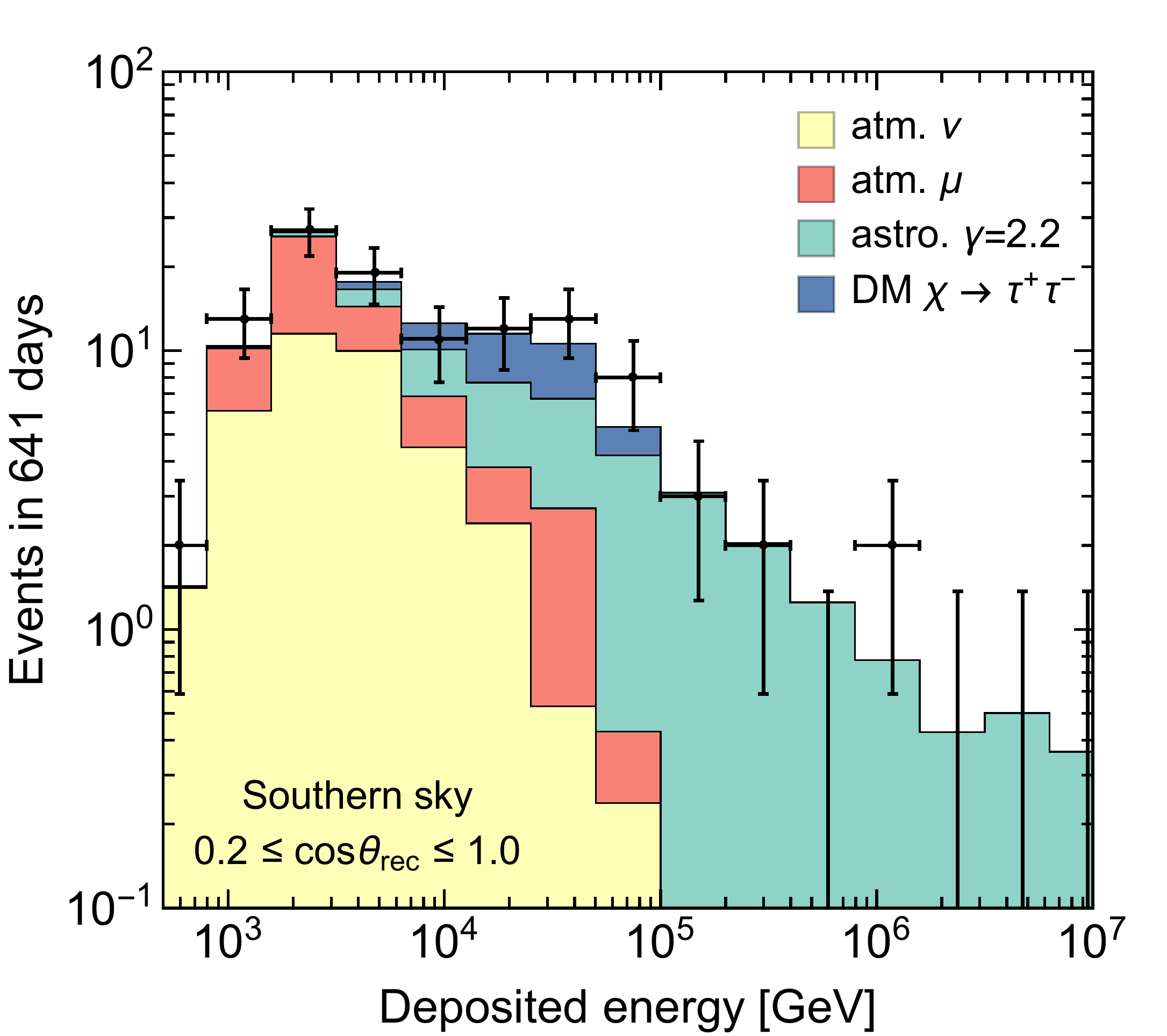}
\hskip2.mm
\includegraphics[width=0.42\textwidth]{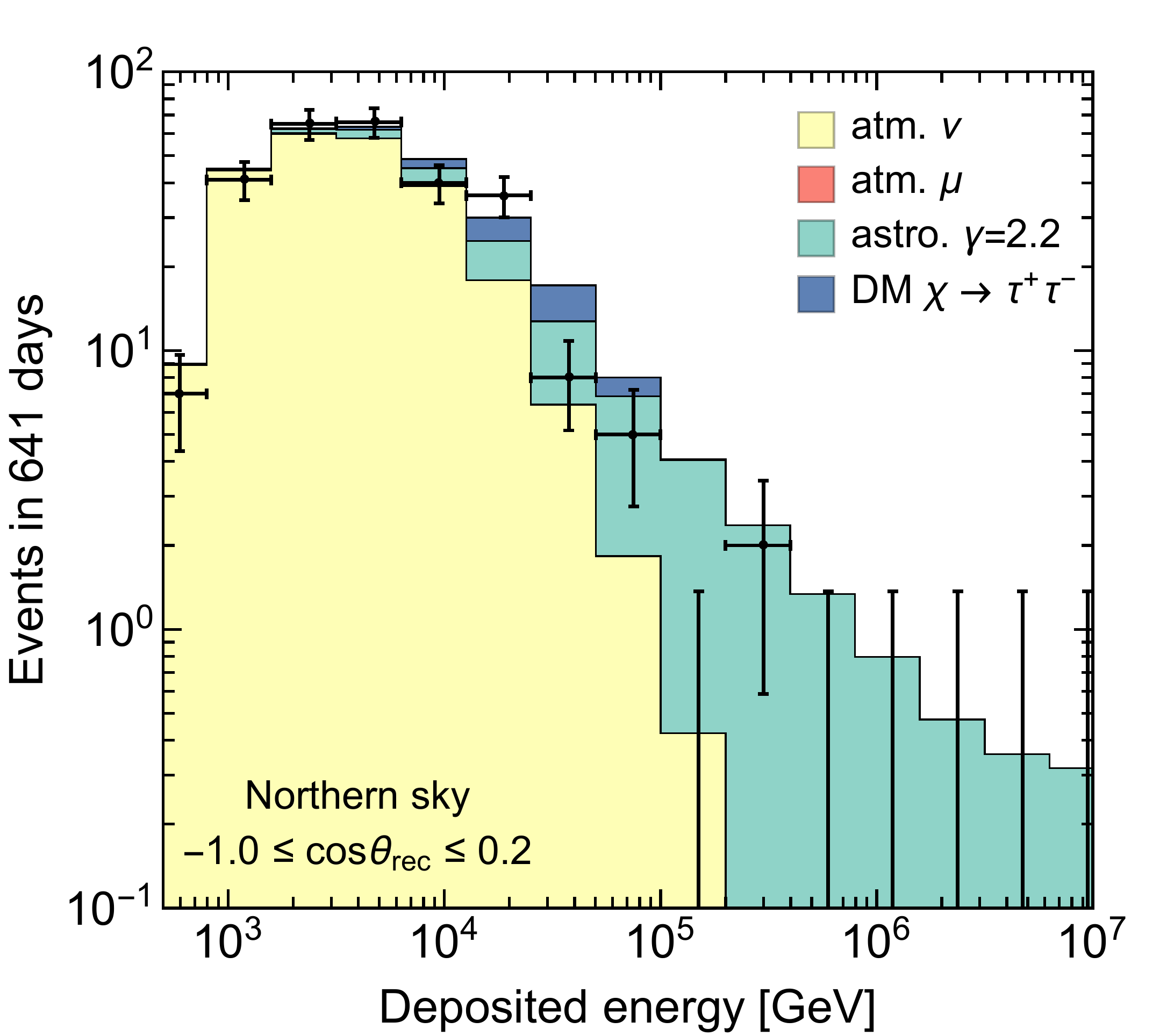}
\end{center}
\caption{\label{fig:events}Numbers of neutrinos events as a function of the neutrino deposited energy after 641 days of data-taking, in the southern (left panels) and northern (right panels) hemispheres. The astrophysical component in green colour is obtained by a power-law behavior with a spectral index 2.0 (upper panels) and 2.2 (lower panels). The DM contribution in blue colour refers to the case of decaying DM model $\chi\to\tau^+\tau^-$ with $m_{\rm DM} = 140$~TeV and $\tau_{\rm DM} = 6\times 10^{27}$~sec ($\tau_{\rm DM} = 9\times 10^{27}$~sec) for $\gamma=2.0$ ($\gamma=2.2$). }
\end{figure}

In this paper, we focused on the 2-years Medium Energy Starting Events (MESE) of IceCube experiment. This set of events exhibits an excess at low energy (10--100~TeV) with respect a single astrophysical power-law, in both southern and northern hemispheres. The maximum {\it local} statistical significance of such an excess is about $1.5\sigma$ in case of the IceCube best-fit spectral index 2.46. The significance increases towards $2.3\sigma$ ($1.9\sigma$) once a harder astrophysical neutrino spectrum with a spectral index equal to 2.0 (2.2) is considered. Indeed, multi-messenger studies and the recent IceCube analysis on the 6-years up-going muon neutrinos suggest a spectral index smaller than 2.2. In the present analysis, we have performed a study to determine the statistical relevance of a scenario where such an excess is due to Dark Matter. In particular, in addition to the atmospheric background (neutrinos and penetrating muons), we consider a two-components neutrino flux, one given by an astrophysical power-law and the other coming from DM particles. The proposed two-components scenario is depicted in Fig.~\ref{fig:events}, where the DM contribution to the observed number of neutrino events is clearly shown in southern and northern hemispheres. 

We have analyzed the case of decaying and annihilating DM particles for quarks and leptons final states with two different DM halo density profiles. The analyses have taken into account the distinction between southern and northern hemispheres in angular coordinates. Remarkably, under the prior of a spectral index in the interval $\left[2.0,2.2\right]$, the statistical relevance of a two-components scenario with respect to a single power-law ranges between about 2$\sigma$--4$\sigma$. By comparing Fig.s~\ref{fig:dec_quark},~\ref{fig:dec_lepton}~and~\ref{fig:dec_A4} with Fig.~\ref{fig:ann}, the decaying models result to be less constrained than the annihilating ones. Moreover, the leptonic final states are favoured with respect to the hadronic ones, since they provide a smaller contribution to the Fermi-LAT gamma-rays. This implies that in case of leptophilic DM models the diffuse gamma-rays flux is mainly explained in terms of astrophysical sources. It is worth observing that our conclusions for annihilating dark matter were derived assuming a particular clumpiness parametrization. However, even though a change in this parametrization is going to slightly affect the allowed range for $\left< \sigma v \right>$, from a qualitatively point of view the results remain unchanged.

\section*{\bf Acknowledgments}

We thank Kohta Murase for the useful comments. The authors acknowledge support by the Istituto Nazionale di Fisica Nucleare I.S. TASP and the PRIN 2012 ``Theoretical Astroparticle Physics" of the Italian Ministero dell'Istruzione, Universit\`a e Ricerca.

\end{document}